\documentclass{aastex62}

\usepackage{graphicx}
\usepackage{mwe}
\usepackage{hyperref}
\usepackage{mathtools}
\usepackage{color}
\usepackage{multirow}

\graphicspath{ }

\newcommand{\link}[1]{{\small{\texttt{\underline{#1}}}}}
\newcommand{\vplanet}{\texttt{VPLanet}\xspace}
\newcommand{\poise}{\texttt{POISE}\xspace}

\def\etal{{\it et al.\thinspace}}
\def\eg{{\it e.g.,}\xspace}

\def\ie{{\it i.e.,}\xspace}
\def\cf{{\it c.f.,}\xspace}
\def\gsim{~\rlap{$>$}{\lower 1.0ex\hbox{$\sim$}}}
\def\lsim{~\rlap{$<$}{\lower 1.0ex\hbox{$\sim$}}}
\def\co2{{CO$_2$}}

\shorttitle{Ice Coverage on Earth-Like Exoplanets}
\shortauthors{C. Wilhelm \etal}

\begin{document} 
    
\title{The Ice Coverage of Earth-like Planets Orbiting FGK Stars}
    
\author{Caitlyn Wilhelm}
\affil{Department of Astronomy, University of Washington, Seattle, WA 98195-1580, USA}
\affil{NASA Virtual Planetary Laboratory, USA}
\email{cwilhelm@uw.edu}

\author{Rory Barnes}
\affil{Department of Astronomy, University of Washington, Seattle, WA 98195-1580, USA}
\affil{NASA Virtual Planetary Laboratory, USA}

\author{Russell Deitrick}
\affil{NASA Virtual Planetary Laboratory, USA}
\affil{Center for Space and Habitability, University of Bern, Gesellschaftsstrasse 6, CH-3012, Bern, Switzerland}

\author{Rachel Mellman}
\affil{Department of Astronomy, University of Washington, Seattle, WA 98195-1580, USA}
    
\begin{abstract}
The photometric and spectroscopic signatures of habitable planets orbiting FGK stars may be modulated by surface ice coverage. To estimate its frequency and locations, we simulated the climates of hypothetical planets with a 1D energy balance model and assumed that the planets possess properties similar to modern Earth (mass, geography, atmosphere). We first simulated planets with fixed rotational axes and circular orbits, finding that the vast majority ($\gsim 90$\%) of planets with habitable surfaces are free of ice. For planets with partial ice coverage, the parameter space for ice caps (interannual ice located at the poles) is about as large as that for ``ice belts'' (interannual ice located at the equator), but belts only persist on land. We then performed simulations that mimicked perturbations from other planets by forcing sinusoidal orbital and rotational oscillations over a range of frequencies and amplitudes. We assume initially ice-free surfaces and set the initial eccentricity distribution to mirror known exoplanets, while the initial obliquity distribution matches planet formation predictions, \ie favoring 90$^\circ$. For these dynamic cases, we find again that $\sim$90\% of habitable planets are free of surface ice for a range of assumptions for ice's albedo. Planets orbiting F dwarfs are three times as likely to have ice caps than belts, but for planets orbiting K and G dwarfs ice belts are twice as likely as caps. In some cases, a planet's surface ice can cycle between the equatorial and polar regions. Future direct imaging surveys of habitable planets may be able to test these predictions.
\end{abstract}

\section{Introduction}
The discovery of small planets in the habitable zone \citep[HZ;][]{Kasting93, Kopparapu13} of main-sequence stars has raised the possibility that we may soon detect a planet with Earth-like surface conditions. Moreover, data from future 10 m class space telescopes and 30 m class ground telescopes may even  reveal surface inhomogeneities such as land, water, and ice \citep[\eg][]{Cowan09,Luger21}. On Earth, the atmosphere reflects a significant amount of solar photons, but ice at the poles contributes about half the total albedo at those latitudes \citep{DonohoeBattisti11}. Thus, if a similar pattern exists for exoplanets, we may be able to identify the locations of ice sheets on their surfaces. To help determine whether planets' ice and land distributions can be observed, we have performed a series of climate simulations designed to predict the distribution of surface ice on Earth-like exoplanets.

To perform these simulations we used a 1-D energy balance model (EBM) that has been calibrated to Earth's climate over 1 year, as well as its Milankovitch cycles \citep{Deitrick18b,Barnes20}. This model is much simpler than general circulation models (GCMs) \citep[\eg][]{Way2017,Wolf20}, but this simplicity enables rapid climate calculations that in turn enable simulations of a large number of hypothetical planets. While individual simulations may be imprecise, EBMs can identify general trends, such as the fraction of planets with or without ice coverage, as well as the location of those icy regions.

Using an Earth-calibrated EBM has the advantage of validation, but it  cannot simulate the full range of possibilities for habitable exoplanets. Thus, we must limit our study to approximately ``Earth-like'' planets, by which we mean planets with masses, rotation periods, geographies, and atmospheres that are very similar to modern Earth. Nonetheless, this approach can provide a baseline level of expectations for the range of habitable climates and ice sheet locations that may be observed with future ground- and space-based observatories. Specifically, we seek to constrain the frequency and locations of highly reflective surface ice that may induce photometric variations over an orbital period. While numerous investigations have explored individual cases or a limited range of parameter space \citep[\eg][]{Williams97,Spiegel09,Dressing10,Spiegel10,Forgan16,HaqqMisra16,Deitrick18b}, no study that we are aware of has systematically explored the obliquity-eccentricity parameter space of Earth's climate with an EBM.  We performed more than 200,000 simulations over a wide range of assumptions to calculate the probability of different ice states, providing a first-order glimpse into the types of ice coverage on habitable and maybe even inhabited worlds.

One aspect of ice coverage that has not been thoroughly explored is the possibility of ice belts along planetary equators. Previous research has found that planets with obliquity $\varepsilon \gsim~ 54^\circ$ may form an equatorial ice belt since the orbit-averaged instellation is lower at the equator than the poles \citep{WilliamsPollard03}. That study and others \citep{Kilic17,Kilic18} found that ice belts can form and persist with 3D GCMs, but \cite{Ferreira14} found that ocean heat transport could prevent the formation of an ice belt. \cite{Rose2017} used an EBM to map the stable regions of ice formation at different obliquities, assuming a planet on a circular orbit around a Sun-like star. Previous studies also only considered Sun-like stars, so the likelihood of ice belt formation on planets with different stellar hosts is largely unknown. Here we expand the study of ice belts by considering their stability under nonsolar irradiation, as well as orbital and rotational oscillations.

We considered two broad categories of simulations that we call ``static'' and ``dynamic.'' For static cases we simulated planets with no orbital or rotational variations and calculated the equilibrium climate. For dynamic cases, we include idealized variations of eccentricity $e$ and obliquity $\varepsilon$ to examine how perturbations affect the location and longevity of surface ice. For both categories, we classify planets into five climate states: moist greenhouse, ice free, ice caps, ice belt, or ``snowball'' (total glaciation). For ice cap and belt planets we also calculate the height of the ice sheet as a function of latitude and subdivide those categories based on the presence of surface ice on land and/or seas.

We consider planets orbiting FGK stars on the main sequence, but not M dwarf systems. Planets orbiting M dwarfs are likely to be synchronous rotators \citep{Dole64,Kasting93,Barnes17} and EBMs do not accurately capture the atmospheric properties of these worlds. Climates of such planets should be treated with GCMs \citep[\eg][]{Joshi97,Yang13,delGenio19}. In the next section we present our methodology, followed by our results, and finally we provide our conclusions. Note that for each figure we include a link to a GitHub website that contains the input files and plotting scripts to generate the figure, \eg \href{https://github.com/caitlyn-wilhelm/IceCoverage/}{\link{IceCoverage}}.

\section{Methods}
        
\subsection{Climate Model}
We briefly describe our climate model here and refer the reader to \cite{Deitrick18b} and \cite{Barnes20} for comprehensive descriptions. We model our hypothetical systems with \vplanet \citep[][]{Barnes20}\footnote{Publicly available at https://github.com/VirtualPlanetaryLaboratory/vplanet.}, which includes \poise (``Planetary Orbit-Influenced Simple EBM''), a one-dimensional seasonal EBM that reproduces Earth's annual climate and its Milankovitch cycles \citep[see][]{NorthCoakley79, HuybersTziperman08}. Though the model lacks a true longitudinal dimension, each latitude is divided into a land portion and a sea portion with distinct heat capacities and albedos, and heat is allowed to flow between them. Specifically, the heat capacity over land is $1.55 \times 10^7$ J m$^{-2}$ K$^{-1}$ and $3.1 \times 10^8$  J m$^{-2}$ K$^{-1}$, the latter of which corresponds to an ocean mixed layer depth of 70 m. We employ the ``outgoing longwave radiation model'' of \cite{NorthCoakley79} and assume the latitudinal diffusion parameter is constant regardless of other planetary/orbital properties. We further assume a uniform 75\%/25\% split between sea and land, respectively. Ice accumulates on land at a constant rate when temperatures are below 0$^{\circ}$ C, while melting/ablation occurs when ice is present and temperatures are above 0$^{\circ}$ C. Sea ice forms when a latitude's temperature drops below $-2^\circ$ C, and melts when higher. To account for ice sheet flow on land, bedrock depression, lithospheric rebound, and ice sheet height, \vplanet employs the formulations from \cite{HuybersTziperman08}. The bedrock depresses and rebounds locally in response to the changing weight of ice above, always seeking isostatic equilibrium. \poise is thus a self-consistent model for the growth and retreat of ice on land and seas due to stellar radiative forcing and angular momentum evolution. 

We also make various assumptions about the reflectivity of water ice, such as its dependence on stellar effective temperature $T_{eff}$. On Earth, snow and ice have a wide range of albedos \citep[see \eg][]{Warren19}, but our climate model lacks the fidelity to model variations in ice albedo due to age, depth, etc. Water ice's absorption depends on wavelength, so its albedo also depends on stellar type \citep{JoshiHaberle12,Shields13}. To further simplify our model, we do not consider albedo variations due to cloud cover, which are important for Earth \citep{DonohoeBattisti11}. Instead we follow the approach of \cite{Deitrick18b}, who calibrated the ice and non-ice surface albedos to reproduce Earth's climate with the EBM, \ie the albedo effects of the atmosphere are implicitly included in the choice of surface albedo. This approach follows previous studies \citep{Shields13,Palubski20} that assumed a constant albedo for ice on a particular planet as a function of stellar effective temperature. Nonetheless, the simplifications in EBM treatments of climate imply our results only provide first-order estimates of the distribution of ice on potentially habitable exoplanets.

\begin{figure}[ht]
\centering
\includegraphics[width=0.9\textwidth]{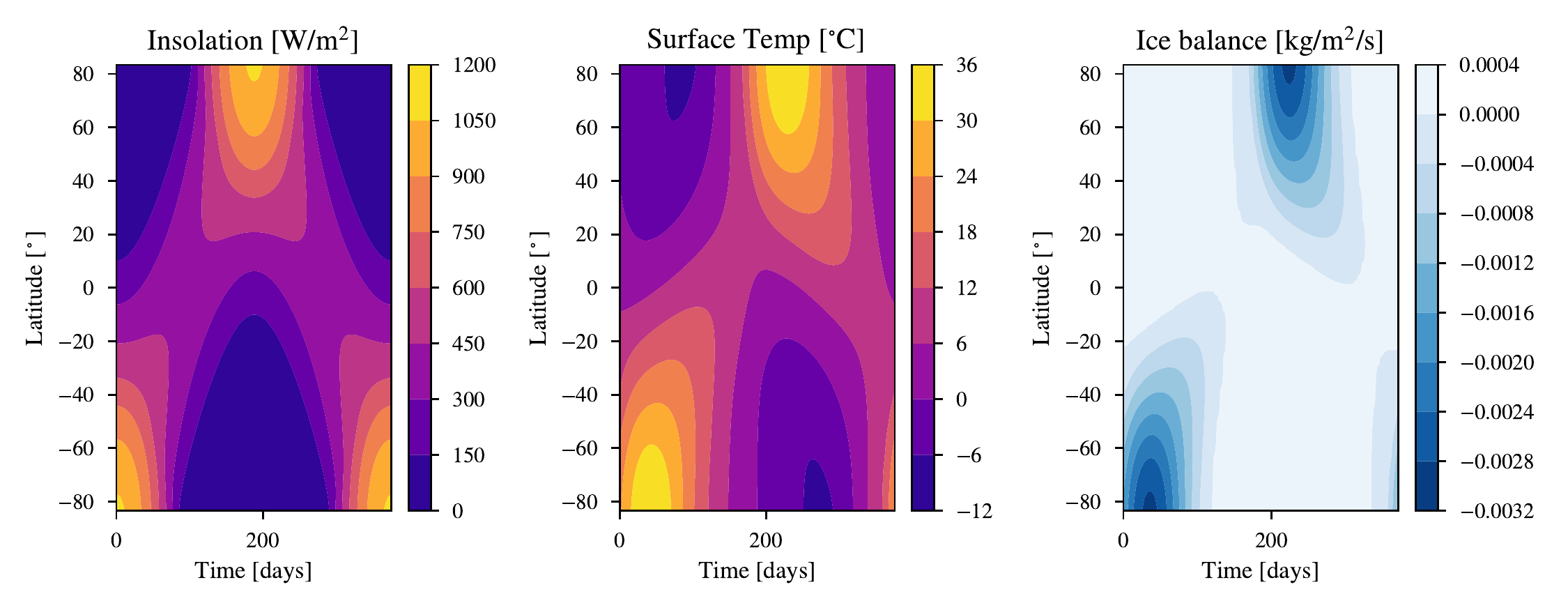}
\caption{Annual climate of an Earth-like planet if its obliquity is 55$^\circ$ and its insolation is 96\% of the solar constant. {\it Left:} The incident solar radiation as a function of day and latitude. {\it Middle:} Surface temperature. {\it Right:} Ice formation rate on land. Positive values indicate ice accumulation, negative ice melting/ablation. \href{https://github.com/VirtualPlanetaryLaboratory/vplanet/tree/main/examples/IceBelts}{\link{VPLanet:examples/IceBelts}}
}
\label{fig:Rose_example}
\end{figure}

\begin{figure}[!h]
\centering
\includegraphics[width=0.7\textwidth]{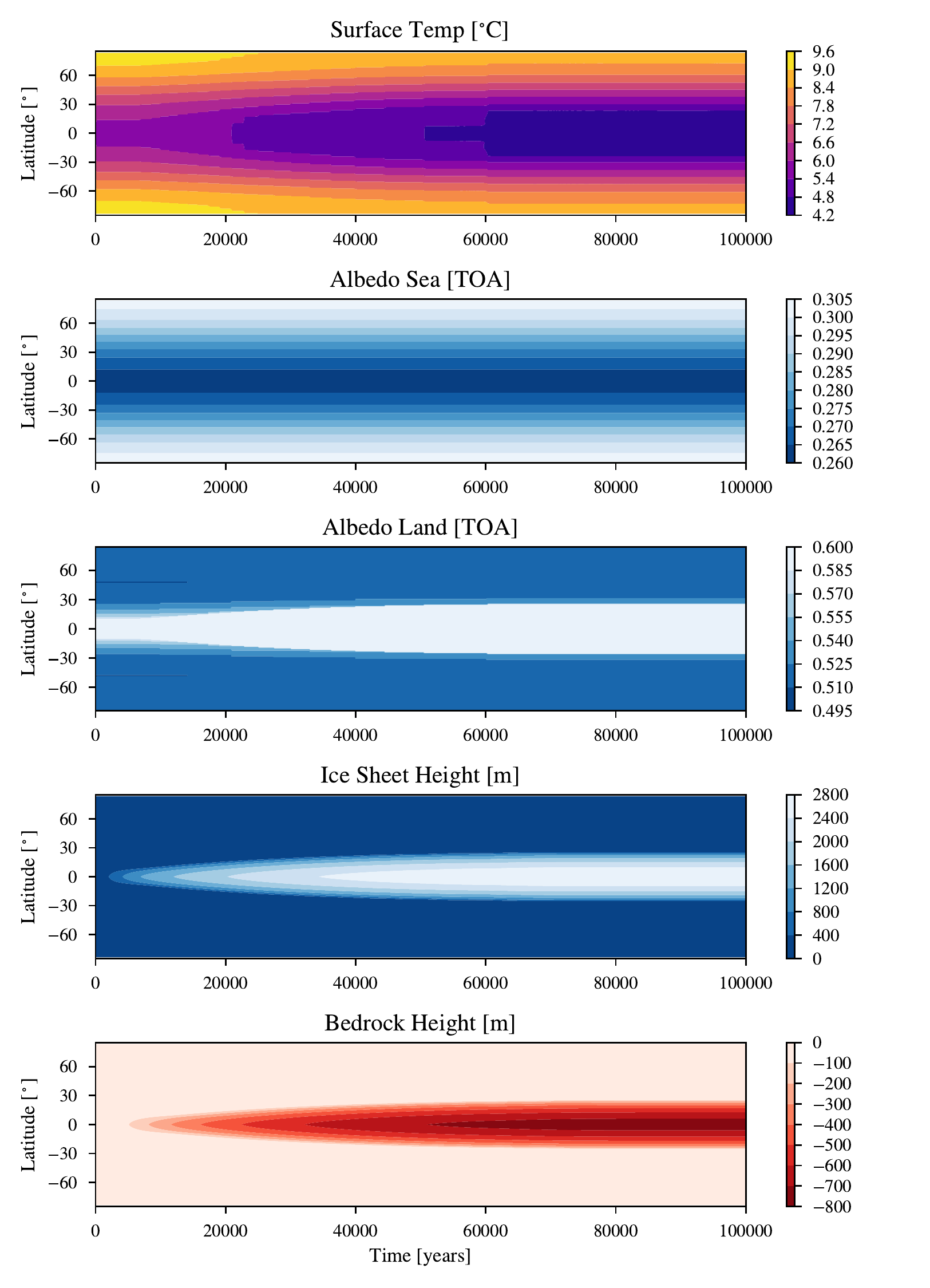}
\caption{Formation of an ice belt on the planet shown in Fig.~\ref{fig:Rose_example}. {\it Top:} Surface temperature. Note that the values are averaged over the orbit and over land and sea. Temperatures over land are, in fact, below freezing for much of the year. {\it Next:} Albedo of the sea at the top of the atmosphere (TOA). {\it Next:} Albedo of the land at the top of the atmosphere (TOA). {\it Next:} Height of the ice sheet on land. {\it Bottom:} Depression of the lithosphere due to ice mass loading. Note that in this case sea ice does not form. The open water at the equator maintains an annually averaged temperature above $0^{\circ}$ C even though ice sheets persist throughout the year on land.  \href{https://github.com/VirtualPlanetaryLaboratory/vplanet/tree/main/examples/IceBelts}{\link{VPLanet:examples/IceBelts}}
}
\label{fig:Rose_evolve}
\end{figure}

Before proceeding, we first evaluate \poise in the high obliquity regime by reproducing one of the results of \cite{Rose2017} in Fig.~\ref{fig:Rose_example}. Specifically, we compare \poise's predictions at $\varepsilon = 55^\circ$ for which Rose et al. found an ice belt formed. \poise similarly forms an ice belt, but only on land. For completeness, in Fig.~\ref{fig:Rose_evolve}, we show the formation of the ice belt, along with the evolution of the climate, albedo, and lithosphere for this case, whose equatorial sea does not freeze. This figure shows that equatorial land freezes over almost immediately and the ice belt requires about 50,000 years to reach equilibrium. The sea albedo remains constant for the duration of the simulation, however, because it never freezes over. Note that the albedos in Fig. \ref{fig:Rose_evolve} are the top-of-atmosphere (TOA) albedo, which incorporates the zenith angle effect of the atmosphere and an approximation for cloud coverage \citep[see][for an exact description]{Deitrick18b}.

\subsection{Obliquity and Eccentricity Oscillations}

To estimate the frequency and locations of surface ice on Earth-like planets, we must include realistic assumptions for $e$, $\varepsilon$, and the precession angle $\psi$ (\ie the azimuthal angle of the rotation axis). To evaluate the roles of these parameters, we assume that $e$ follows the distribution of non-tidally-evolved exoplanets, which peaks slightly above 0 and has a long tail to $\sim 0.9$ \citep[see \eg][]{Barnes15_tides}. The initial $\varepsilon$ and $\psi$ of planets are likely determined by the last major impact event of planet formation, which, in aggregate, produce an isotropic distribution of the rotational angular momentum \citep{LinIda97,Chambers99,MiguelBrunini10}. This results in a non-uniform distribution of obliquities that peaks at 90$^{\circ}$. Note that the propensity of Solar System terrestrial planets to currently possess $\varepsilon < 25^\circ$ does not imply exoplanets will be similar. Mercury and Venus have both experienced strong tidal evolution, which damps obliquities toward 0 \citep{GoldSoter69,Hut81,Heller11}. Mars' obliquity has evolved chaotically and with large amplitude since at least the Noachian \citep{Laskar1993,BrasserWalsh11}, indicating that its current low value is just a coincidence. Thus, we discard the Solar System's current $\varepsilon$ distribution as an appropriate guide for the initial conditions of the galactic population of terrestrial exoplanets.

In multiplanet systems, e, $\varepsilon$, $\psi$ evolve with time due to gravitational interactions with other planets and the host star \citep{Kinoshita1975,Laskar1993,Deitrick18a}. The architectures of planetary systems are still relatively mysterious, so the range of possibilities creates a very large parameter space to explore. \cite{Deitrick18b} mitigated this complexity by focusing on a few specific orbital configurations and explored how the climates could evolve. Here we take a complimentary approach: rather than build synthetic planetary systems, we impose a range of oscillations on $e$ and $\varepsilon$ to mimic these interactions. In real systems, multiple frequencies exist, but we assume that our approximation is commensurate with the simplicity of our climate model -- we seek only a first order estimate of the distribution of surface ice. The precession of the rotational axis is due to the stellar torque and depends on $e$, $\varepsilon$, and the shape of the planet. We therefore calculated axial precession explicitly assuming the planets maintain the same rotational flattening as Earth.

We imposed sinusoidal oscillations of $e$ and $\varepsilon$ with amplitudes and frequencies that are similar to those in our Solar System \citep[see \eg][]{Laskar88} and exoplanet systems \citep[see \eg][]{BQ04}. Thus, obliquity evolves according to 
\begin{equation}
\varepsilon(t) = \frac{A_\varepsilon}{2}\sin \left(\frac{2\pi t}{P_\varepsilon}\right) + \varepsilon_{0},
\end{equation}
where $t$ is time, $\varepsilon(t)$ is the obliquity as a function of time, $A_\varepsilon$ is the obliquity amplitude, $P_\varepsilon$ is the obliquity period, and $\varepsilon_{0}$ is the initial obliquity. Similarly, eccentricity evolves as 
\begin{equation}
e(t) = \frac{A_e}{2}\sin \left(\frac{2\pi t}{P_e}\right) + e_{0},
\end{equation}
where $e_0$ is the initial eccentricity, $A_e$ is the eccentricity amplitude, and $P_e$ is the eccentricity period. 
 
For evolving systems we also include the precession of the rotational axis due to the stellar torque \citep{Kinoshita1975,Kinoshita1977,Deitrick18a}. We use the standard convention for $\psi$ as shown in Fig.~1 in \cite{Deitrick18a}. With these assumptions, the evolution of $\psi$ is described as
\begin{equation}
\psi(t) = \frac{3 G M_* \epsilon_D (1-e^2)^{-3/2} \cos{\varepsilon}}{2 a^3 \nu}  t + \psi_0,
\label{eq:solartorque}
\end{equation}
where $G$ is Newton's gravitational constant, $M_*$ is the stellar mass, $a$ is the semi-major axis of the planet, $\nu$ is the planet's rotational frequency, $\epsilon_D$ is the planet's dynamical ellipticity, and $\psi_0$ is the initial value of the precession angle. For the planets we simulate, we assume Earth's value for $\epsilon_D$ (0.00328) and a rotation period of 1 day.  Thus, as $e$ and $\varepsilon$ evolve, $\psi$ responds self-consistently. In practice, for static cases we set $\epsilon_D = 0$ to suppress axial precession. Note that the EBM does not explicitly depend on rotation period.
 
\subsection{Initial Conditions and Parameter Space}
We performed two types of experiments: non-evolving (static) planets and evolving planets. We always assume an ice-free initial state (``warm start'') for evolving cases, but explore both warm start and cold start conditions  (total glaciation) for static cases. Warm start conditions have a global average temperature $+15^{\circ}$ C; cold start conditions have a global average temperature $-40^{\circ}$ C and a thin layer of ice on land. Testing revealed no notable difference in outcomes as a function of initial ice height. 

We simulated static cases for 1000 years and evolving cases for at least $10^6$ years, timescales that we find are sufficiently long for the climates to stabilize. 
For the static cases, we follow \cite{Shields13} for luminosity $L$ and effective temperature $T_{eff}$ values for the different spectral types and the associated ice albedo as a function of stellar effective temperature. Within each spectral type, we hold the ice albedo constant for the static planet; we explore the effect of variations in the ice albedo in our evolving cases. For each spectral class, we explored $\varepsilon$ between 0 and 90$^\circ$ and $e$ between 0 and 0.5. We assume the climate states are symmetric about $\varepsilon = 90^\circ$, the natural expectation for fast rotators. We set $\varpi = \psi = 0$, which means that the equinoxes occur at peri- and apocenter. We then searched for the instellation $S$ (normalized to the solar constant) and $\varepsilon$ limits that defined ice free, ice belt, ice cap, and snowball regions. We refer to these 4 categories as the ``ice state'' of a planet. Model parameter ranges are summarized in Table \ref{tab:static}, where  $L_{\odot}$ is the solar luminosity and $M_{\odot}$ is the solar mass.
            
\begin{table}[h]
\centering
\caption{Model Settings for Static Simulations}
\begin{tabular}{lccc}
\hline \hline
Parameter                & K                  & G                  & F           \\ \hline
Luminosity ($L_{\odot}$) & 0.34               & 1                  & 3.46        \\
$T_{eff}$ (K)            & 5084               & 5887               & 6594        \\
$M_*$ ($M_{\odot}$)      & 0.82               & 1                  & 1.48        \\
$a$ (Warm Start) (au)      & [0.58, 0.63]       & [1, 1.07]          & [1.78, 1.94]\\
$a$ (Cold Start) (au)      & [0.49, 0.58]    & [0.83,0.97]        & [1.45, 1.8] \\
Ice albedo               & 0.55               & 0.6                & 0.63        \\
\hline
\end{tabular}
\label{tab:static}
\end{table}

For the evolving cases, we performed 5 sets of 10,000 simulations for each spectral type. Each set makes different assumptions regarding albedo and the dynamical properties of the planet's orbit, allowing an estimate of the robustness of the results. Since little is known about the true behavior of habitable exoplanets, these assumptions are arbitrary, but are chosen to reflect reasonable expectations of their properties. We refer to these different sets of simulations as Cases A--E and detail their settings in Table \ref{tab:evolve}. We varied parameters either uniformly or normally within the limits shown in Table \ref{tab:evolve}, where $S_\oplus$ is the modern Earth's solar constant (1361 W/m$^2$). The columns for uniformly sampled parameters contain the minimum and maximum while the columns for normally distributed parameters show the mean and standard deviation of the distribution.
The ranges for $a$ as a function of host star spectral type are determined from the results of the static cases presented in $\S$3 and assumptions about the onset of the moist greenhouse (described shortly).

\begin{table}[h]
\centering 
\caption{Selected Parameters and Their Distributions for the Evolving Cases}
\begin{tabular}{ccccccccc} \hline \hline
Case & Spectral & $a^{uniform}$ & $P_e^{uniform}$ & $P_e^{normal}$ & $P_\varepsilon^{uniform}$ & $P_{\varepsilon}^{normal}$ & $T$ & Albedo\\
 & Type & (au) & (yr) & (yr) & (yr) & (yr) & (days) &  \\ \hline
\multirow{3}{*}{A} & K & {[}0.488,0.63{]} & {[}1000,$10^5${]} & N/A & {[}1000,$10^5${]} & N/A & {[}1,1{]} & {[}0.55,0.55{]} \\
 & G & {[}0.83,1.07{]} & {[}1000,$10^5${]} & N/A & {[}1000,$10^5${]} & N/A & {[}1,1{]} & {[}0.6,0.6{]} \\
 & F & {[}1.55,1.94{]} & {[}1000,$10^5${]} & N/A & {[}1000,$10^5${]} & N/A & {[}1,1{]} & {[}0.63,0.63{]} \\ \hline
\multirow{3}{*}{B} & K & {[}0.488,0.63{]} & {[}1000,$10^5${]} & N/A & {[}1000,$10^5${]} & N/A & {[}1,1{]} & {[}0.5,0.6{]} \\
 & G & {[}0.83,1.07{]} & {[}1000,$10^5${]} & N/A & {[}1000,$10^5${]} & N/A & {[}1,1{]} & {[}0.55,0.65{]} \\
 & F & {[}1.55,1.94{]} & {[}1000,$10^5${]} & N/A & {[}1000,$10^5${]} & N/A & {[}1,1{]} & {[}0.58,0.68{]} \\ \hline
\multirow{3}{*}{C} & K & {[}0.488,0.63{]} & {[}$5\times{10^4}$,$5\times{10^5}${]} & N/A & {[}$5\times{10^4}$,$5\times{10^5}${]} & N/A & {[}1,1{]} & {[}0.5,0.75{]} \\
\multicolumn{1}{l}{} & G & {[}0.83,1.07{]} & {[}$5\times{10^4}$,$5\times{10^5}${]} & N/A & {[}$5\times{10^4}$,$5\times{10^5}${]} & N/A & {[}1,1{]} & {[}0.5,0.75{]} \\
\multicolumn{1}{l}{} & F & {[}1.55,1.94{]} & {[}$5\times{10^4}$,$5\times{10^5}${]} & N/A & {[}$5\times{10^4}$,$5\times{10^5}${]} & N/A & {[}1,1{]} & {[}0.5,0.75{]} \\ \hline
\multirow{3}{*}{D} & K & {[}0.488,0.63{]} & N/A & {[}$10^4$,25000{]} & N/A & {[}$10^4$,25000{]} & {[}1,1{]} & {[}0.5,0.75{]} \\
 & G & {[}0.83,1.07{]} & N/A & {[}$10^4$,25000{]} & N/A & {[}$10^4$,25000{]} & {[}1,1{]} & {[}0.5,0.75{]} \\
 & F & {[}1.55,1.94{]} & N/A & {[}$10^4$,25000{]} & N/A & {[}$10^4$,25000{]} & {[}1,1{]} & {[}0.5,0.75{]} \\ \hline
\multirow{3}{*}{E} & K & {[}0.488,0.63{]} & {[}1000,$10^5${]} & N/A & {[}1000,$10^5${]} & N/A & {[}0.5,5{]} & {[}0.55,0.55{]} \\
 & G & {[}0.83,1.07{]} & {[}1000,$10^5${]} & N/A & {[}1000,$10^5${]} & N/A & {[}0.5,5{]} & {[}0.6,0.6{]} \\
 & F & {[}1.55,1.94{]} & {[}1000,$10^5${]} & N/A & {[}1000,$10^5${]} & N/A & {[}0.5,5{]} & {[}0.63,0.63{]} \\ \hline
 \label{tab:evolve}
\end{tabular}
\end{table}

The 5 cases are designed to test the role of albedo and external forcing on the climates of planets orbiting stars of different spectral type. In Case A, our baseline model, we assume all albedos have the values shown in Table \ref{tab:static} and the amplitudes and periods of the $e$ and $\varepsilon$ oscillations are chosen to represent typical values for the terrestrial planets in our Solar System. In Case B, we allowed the albedos to take values within $\pm 0.05$ the values (with a uniform distribution) from Table \ref{tab:static}, \ie the general trend in albedo and $T_{eff}$ is preserved, but some variations in ice albedo are permitted. The external forcing oscillations are the same as in Case A. In Case C, the ice albedos for all planets, regardless of host star spectral type, are allowed to vary uniformly between 0.55 and 0.75 to further evaluate variable albedo, while the external forcing periods are shifted to larger values. In Case D, the albedos are the same as Case C, but the external forcing parameters have normally distributed periods. In Case E, the values are the same as Case A but we changed the rotation rate to vary uniformly between 0.5 to 5 days, which changes the evolution of $\psi$, see Eq.~\ref{eq:solartorque}.

Cases A and B are designed to test two different models of ice's albedo, but with the same orbital forcing. Case C removes the assumption of ice albedo correlating with spectral type, so that we are essentially testing the effect of orbital period and long-term forcings. Case D is a variation on Case C that explores the robustness of results against the assumed forcing frequency. Finally Case E explores the robustness of results with different rotational periods. We make two different assumptions regarding the range of oscillation frequencies to explore the relative importance of this parameter. While our choices for the frequency ranges are arbitrary, our selected distributions are plausible and test uniform vs.\ normally distributed frequencies.

As our primary goal is to calculate the ice state fractions for habitable planets, we must define a working definition of the HZ. In $\S$3.1 we find the minimum instellation that permits at least 1 latitude with surface water for static planets and set that $a$ as the outer limit of the HZ. As we assume ice's albedo is a function of $T_{eff}$, this limit is also a function of spectral type. For the inner edge, we assumed the moist greenhouse, in which water penetrates the tropopause and ultimately escapes to space, occurs at $70^\circ$ C in the global mean \citep{Kasting88,WolfToon15,Palubski20}. To pinpoint the location where this occurs, we ran \poise with $\varepsilon = 45^\circ$ at ever decreasing $a$ until the mean global temperature reached $70^\circ$ and defined that value of $S$ to be the inner edge of the HZ. By simulating planets between these limits, we can predict the ice state fractions for habitable planets orbiting FGK stars. Note that we only performed this experiment for warm start planets as we expect planets to be hot after formation, and to enter a ``hot house'' phase immediately after a snowball Earth event on geologically active planets, which is consistent with our assumption that we are simulating Earth-like planets.

Finally, we set the initial values of $e$ and $\varepsilon$ and their amplitudes and periods of oscillation. We employed non-uniform initial distributions of $e_0$ and $\varepsilon_0$ for our hypothetical exoplanets. We assume $e_0$ is distributed normally with a mean at 0.05 and a standard deviation of 0.1 (values less than 0 are thrown out and resampled.) While the initial $e$ distribution of habitable exoplanets is unknown, this choice is similar to the observed distribution of non-tidally-evolved exoplanets \citep[see \eg][]{Barnes15_tides}. We set the initial obliquity distribution to be uniform in cos($\varepsilon_0$) to be consistent with models of planet formation \citep{Chambers99,KokuboIda07,MiguelBrunini10}. For the $e$ and $\varepsilon$ amplitudes, we require the minimum and maximum values to be in the ranges [0,1) and [0,180), respectively. In practice, we first selected $e_0$ and $\varepsilon_0$, determined if it was above or below 0.5 or $90^\circ$, respectively, then set the maximum amplitude accordingly. The initial values of $\psi$ are uniformly sampled in the range [0,360) for all cases. The remaining parameters are the same as in Table \ref{tab:static}. At the end of each simulation we determined if and where surface ice existed on the planet and tabulated the ice fractions and their extents.

\section{Results}
    
\subsection{Static Planets} \label{sec:static}
\begin{figure}[h!]
\centering
\includegraphics[width=0.9\textwidth]{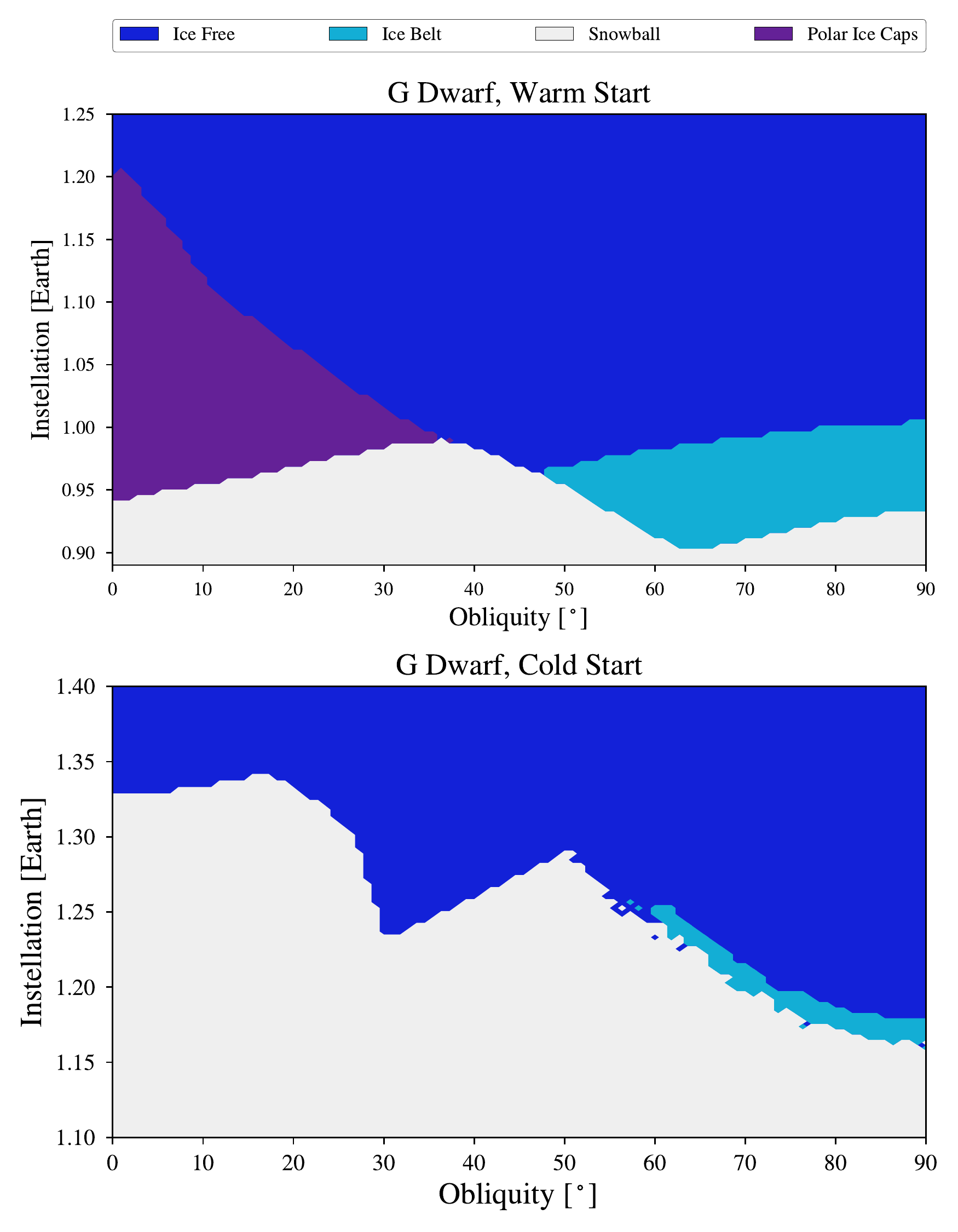}
\caption{{\it Top:} Parameter combinations that permit ice belt formation for a planet orbiting a G dwarf on a circular orbit. The light blue area is where ice belts are stable, dark blue is where no permanent ice is possible, purple is where ice caps are stable, and white is where the planet is completely ice covered. {\it Bottom:} The same Parameters combinations as above, but with a cold start to highlight hysteresis.
\href{https://github.com/caitlyn-wilhelm/IceCoverage/tree/main/StaticCases/GDwarf}{\link{GDwarfStatic}}
}
\label{fig:G_example}
\end{figure}

In this subsection we identify the parameter ranges where ice formation can occur as a function of stellar mass, semi-major axis, eccentricity, and obliquity. For visualization purposes, we calculate the orbit-averaged incident stellar radiation (instellation) relative to Earth: 
\begin{equation}
S = \frac{La^2_{\oplus}}{L_{\odot}a^2\sqrt{(1-e^2)}},
\end{equation}
where $a_{\oplus}$ is the semi-major axis of Earth. In other words, the solar constant corresponds to $S = 1$. In Fig.~\ref{fig:G_example} we show the results for a solar twin host star and a circular orbit for both warm and cold starts. For warm starts, ice caps can form on planets with $\varepsilon < 40^\circ$, while ice belts can form for $\varepsilon > 47^\circ$. Ice caps can occur in the range $0.94 < S < 1.2$, while an ice belt can be present if $0.9 < S < 1$. Earth is located well within the polar cap zone, but if its obliquity were $\gsim 60^\circ$ then it would lie close to the boundary between the ice free and ice belt states. For cold starts, we find, as expected, that ice free states require higher instellation, but also that polar caps are impossible while ice belts are stable.

While polar caps can form on both land and sea, we find that ice belts are confined to land. There is a complex set of factors at work that prevent ice belts over sea. The lower albedo contributes to a higher temperature, while the higher heat capacity dampens the seasonal cycle and allows for warmer winters. Most importantly, sea ice is treated as a thin layer with no thermal inertia, while the thickness of ice on land and its corresponding inertia is explicitly modeled. Accounting for the sea ice thickness may facilitate ice belt formation on the sea. However, sea ice belts will still likely be rarer because the sea doesn't support ice as thick as on land.

The structure in the top panel of Fig. \ref{fig:G_example} is dictated by the ice-albedo feedback and the instellation distribution. At $\varepsilon =0^{\circ}$, a large range of instellations allows for ice caps to form (purple region) because the poles are in permanent darkness, but as the obliquity increases, more stellar flux is received by the poles, lowering the instellation threshold for the ice free state (dark blue). At the same time, the region of stable latitudes for ice cap extent narrows, causing the transition to the snowball state (light gray region) to occur at higher instellation \citep[see][]{Rose2017}. At $\varepsilon \sim 40^{\circ}$, there is no stable location for the ice edge, and only ice free and snowball states exist. The boundary between the two states moves toward lower instellation as the obliquity increases because the (annually averaged) distribution in flux becomes more evenly distributed, decreasing the likelihood that any location on the planet will reach annually averaged temperatures below freezing. Increasing the obliquity further, we reach a state where ice becomes stable at the equator (light blue region). The intensity of summer prevents inter-annual ice from forming at the poles, allowing the temperate ice belt state to push toward lower instellation as obliquity increases, until the trend reverses at $\varepsilon \sim 65^{\circ}$. Here, the decrease in annually-averaged temperatures at the equator begins to dominate over the increase at the poles, resulting in snowball states at higher instellation. A similar trend separates the ice free and ice belt regions: as obliquity increases, the equatorial temperature decreases, allowing ice belts to form at higher instellation.

\begin{figure}[h]
\centering
\includegraphics[width=0.9\textwidth]{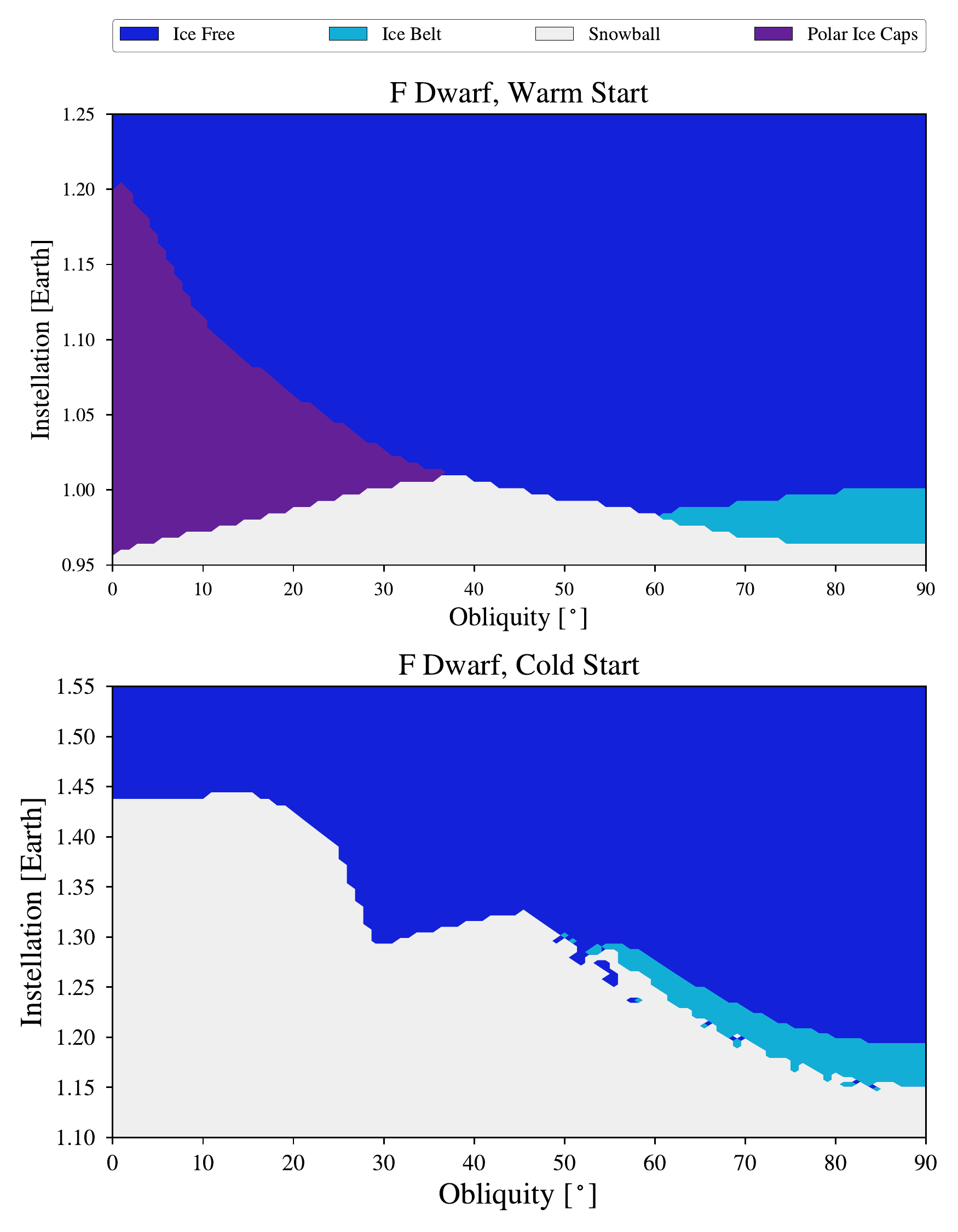}
\caption{Same as Fig.~\ref{fig:G_example}, but for an F dwarf host star. \href{https://github.com/caitlyn-wilhelm/IceCoverage/tree/main/StaticCases/FDwarf}{\link{FDwarfStatic}}
}
\label{fig:F_star}
\end{figure}

\begin{figure}[h]
\centering
\includegraphics[width=0.9\textwidth]{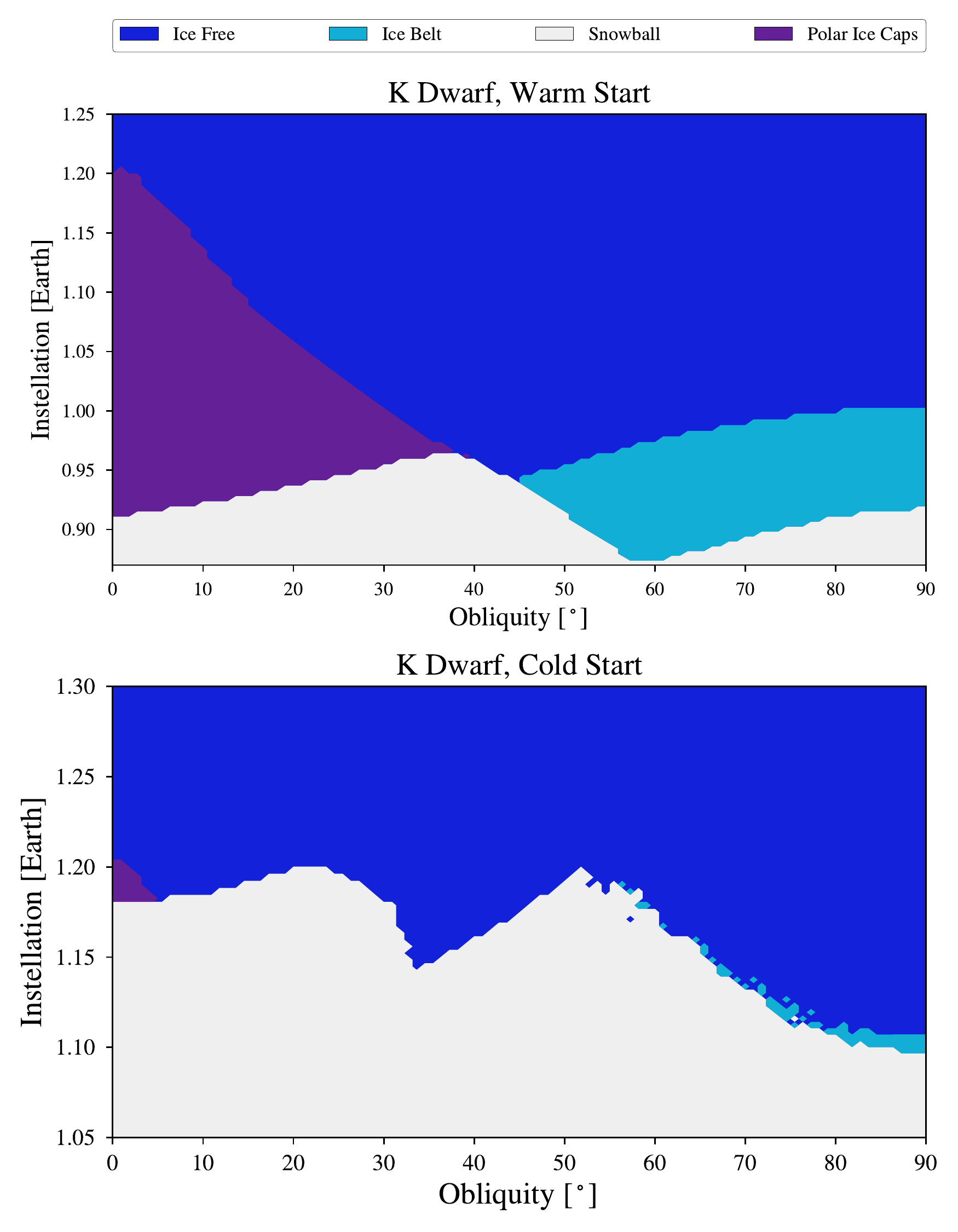}
\caption{Same as Fig.~\ref{fig:G_example}, but for an K dwarf host star. \href{https://github.com/caitlyn-wilhelm/IceCoverage/tree/main/StaticCases/KDwarf}{\link{KDwarfStatic}}
}
\label{fig:K_star}
\end{figure}

The top panels of Figs.~\ref{fig:F_star} and \ref{fig:K_star} show the results for warm starts of F dwarf planets and K dwarf planets, respectively. In each case an instellation of 1 still corresponds to the solar constant. The results are qualitatively similar, but some differences are present in the boundaries between ice states. First we note that polar cap regions are similar for the 3 stellar types. The ice cap region in the F dwarf case has shrunk slightly, compared to the cooler stars, as the threshold for snowball states has moved upward. Second, the ice belt parameter space for F dwarf host stars is smaller than for K and G type host stars, which we attribute to ice's higher albedo pushing the planet into a snowball (see below). Third, the range of obliquities for which there is no stable location for inter-annual ice grows from $40^{\circ}$--$45^{\circ}$ for K dwarfs to $38^\circ$--$60^\circ$ for F dwarf host stars. We attribute this feature to the increasing orbital period (semi-major axis) for habitable planets (see Table \ref{tab:static}): as periods increase, so does the length of summer, which \cite{Deitrick18b} showed is a critical parameter for sustaining long-lived ice sheets.

The first two trends described above are consistent with \cite{Rose2017}, who showed that the stable ice regions diminish as the difference between the ice and land albedo increases, with the effect being more pronounced at high obliquity.
The ice belt state is less stable than the ice cap state for two reasons: (1) the equator receives strong seasonal instellation near the equinoxes, making it easier to tip into the ice-free state; and (2) the large surface area necessarily covered by equatorial sheets makes it easier to tip into the snowball state. Because of this increased susceptibility to instability, the effect of albedo is greater than for ice caps. Consequently, we might expect ice belts to be less common for habitable planets orbiting hotter stars.

The lower panels of Figures \ref{fig:G_example}--\ref{fig:K_star} show the resulting climate state over instellation and obliquity with cold start initial conditions. Hysteresis is substantial for all three host stars and the structure is markedly different from the warm start conditions. Notably, only the K-dwarf planets are able to retain polar caps, but only for a narrow range of parameter space. Otherwise, the low obliquity regions are either ice-free or completely ice covered, and the transition between states occurs at higher instellation than in warm start cases. As expected, because of the change in ice albedo, the cooler the host star is, the lower the threshold in instellation for escaping the snowball state. The paucity of polar ice caps is probably because the instellation threshold for escaping the snowball state is higher than the threshold for triggering the small ice cap instability (wherein a polar cap is unstable). 

The shape of the ice-free/snowball boundary is quite curious and we interpret the snowball boundary as a function of $\varepsilon$ as follows: as $\varepsilon$ increases from 0, the temperature at equatorial latitudes decreases, preventing deglaciation until higher instellation.
At $\varepsilon \sim 30^\circ$, the poles begin to intercept significant starlight and begin to melt, dampening the ice-albedo feedback and pushing the snowball boundary to lower instellation. As $\varepsilon$ continues to increase, the temperature as a function of latitude becomes more isothermal as instellation is spread more evenly on the planet surface and driving the snowball boundary to a local maximum in instellation at $\sim50^\circ$. At this point, summer melting at the poles becomes more intense and allows escape from total glaciation at lower instellation. This effect is maximized at $\varepsilon = 90^\circ$, so this value is the global minimum for the instellation that enables a planet to break out of the snowball state. The boundaries between states at high obliquity appear a bit ``ragged''---this appearance is likely because the tipping points that delineate the regions are sensitive to numerical errors.

\begin{figure}[h]
\centering
\includegraphics[width=\textwidth]{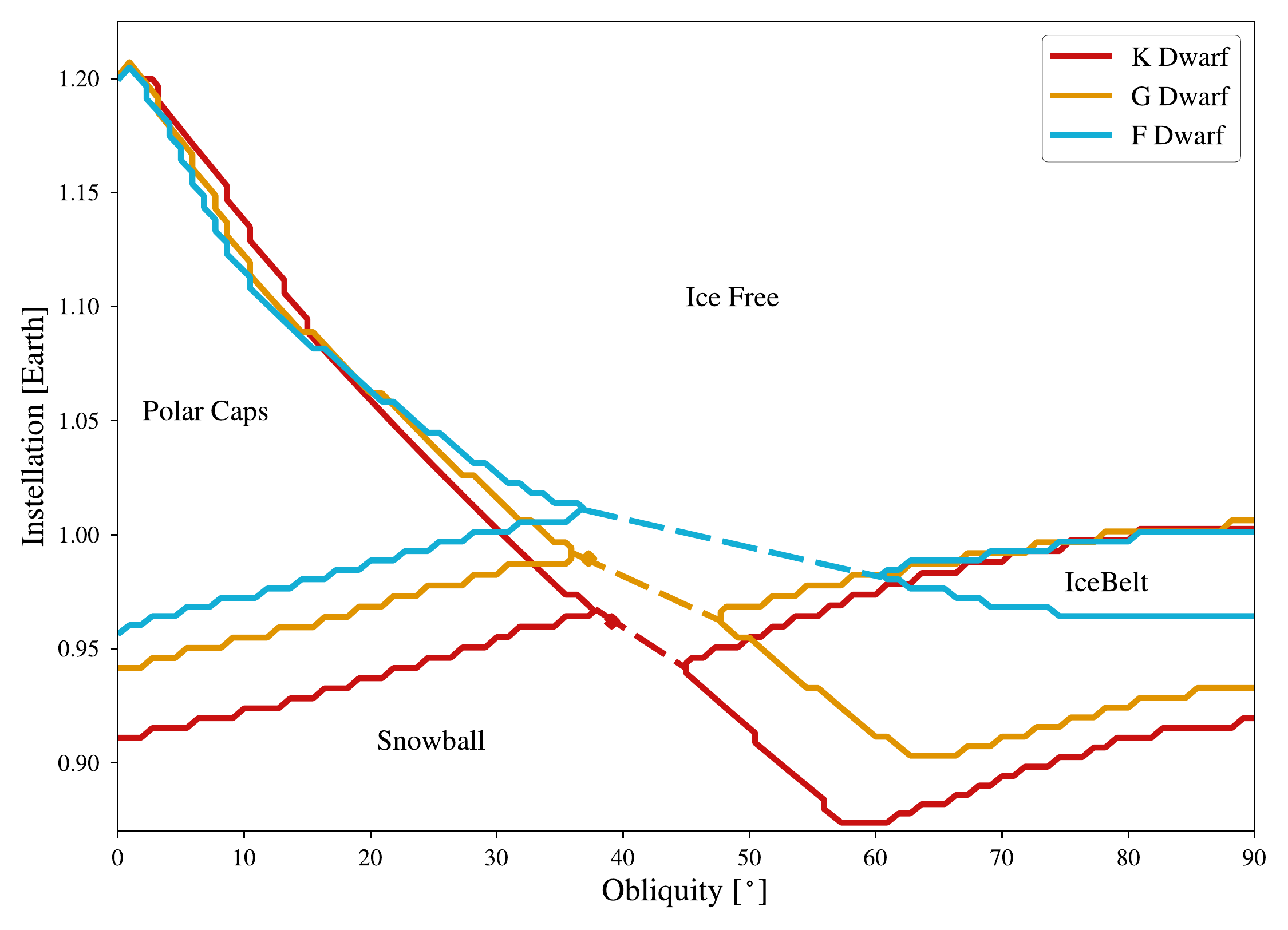}
\caption{Comparison of the ice  caps and belt parameter space ranges for warm start planets on circular orbits of F (light blue), G (orange), and K (red) dwarf stars,  with each region is labeled accordingly. The dashed lines between the cap and belt regions are the approximate location of the boundary between snowball and ice free ice states.
\href{https://github.com/caitlyn-wilhelm/IceCoverage/tree/main/StaticCompare}{\link{StaticCompare}}
}
\label{fig:Startype_Compare}
\end{figure}

In Fig.~\ref{fig:Startype_Compare} we compare the ice sheets for the 3 host star types, with all planets on circular orbits and warm starts, \ie the purple and pale blue regions in the top panels of Figs.~\ref{fig:G_example}--\ref{fig:K_star}. Several features are noteworthy. First, as mentioned above, the ice belt regime for F dwarfs is smaller than for G dwarfs, which is in turn smaller than for K dwarfs. We attribute this trend to the $T_{eff}$--albedo relationship of ice. Note that the ice free/ice belt boundary between all three cases is approximately constant, but that the ice belt/snowball boundary shrinks from K to F dwarf planets -- the ice belts are growing into snowballs for planets orbiting higher mass stars. Third, the minimum instellation for an ice belt occurs at an obliquity of $58^\circ$ for K dwarf planets and an obliquity of $75^\circ$ for F dwarf planets. This can be understood as a consequence of the longer year and the increased ice albedo: winters are longer for the F dwarf planets, allowing ice sheets to grow larger at the poles. At the same time, summers are cooler because of the increased ice albedo. Both effects conspire to make it more difficult for the poles to deglaciate during summer, so that these high obliquity planets are more likely to end up in the snowball state.
Fourth, the minimum obliquity for an ice belt increases from $45^\circ$ to $60^\circ$ for K dwarfs to F dwarfs, because of the effects just described.
Finally, the polar caps region shrinks from K to F dwarf hosts because the higher albedo of ice for F dwarf spectra increases the likelihood for collapse into a snowball.

Next we turn to how eccentricity affects ice coverage. For each host star spectral type, we considered eccentricities up to 0.5 and set $\psi$ to equal the longitude of periastron $\varpi$. The latter choice ensures that the annual instellation distribution is symmetric about the equator and isolates the effects of eccentricity.  We find that ice caps and belts are stable on K dwarfs for $e \le 0.4$, G dwarfs for $e \le 0.3$, and F dwarfs for $e \le 0.1$. We attribute this trend to the orbital period in the HZ: For a planet in the HZ of a K dwarf, the ``summer,'' which occurs globally at pericenter, is much shorter than for a planet in the HZ of an F dwarf. Similarly, the ``winter'' at apocenter is much longer for the F dwarf planets. Thus, the melting at pericenter and ice growth at apocenter is more severe for planets orbiting F dwarfs, increasing the likelihood of both the ice-free and snowball states.

\begin{figure}[h!]
\centering
\includegraphics[width=0.9\textwidth]{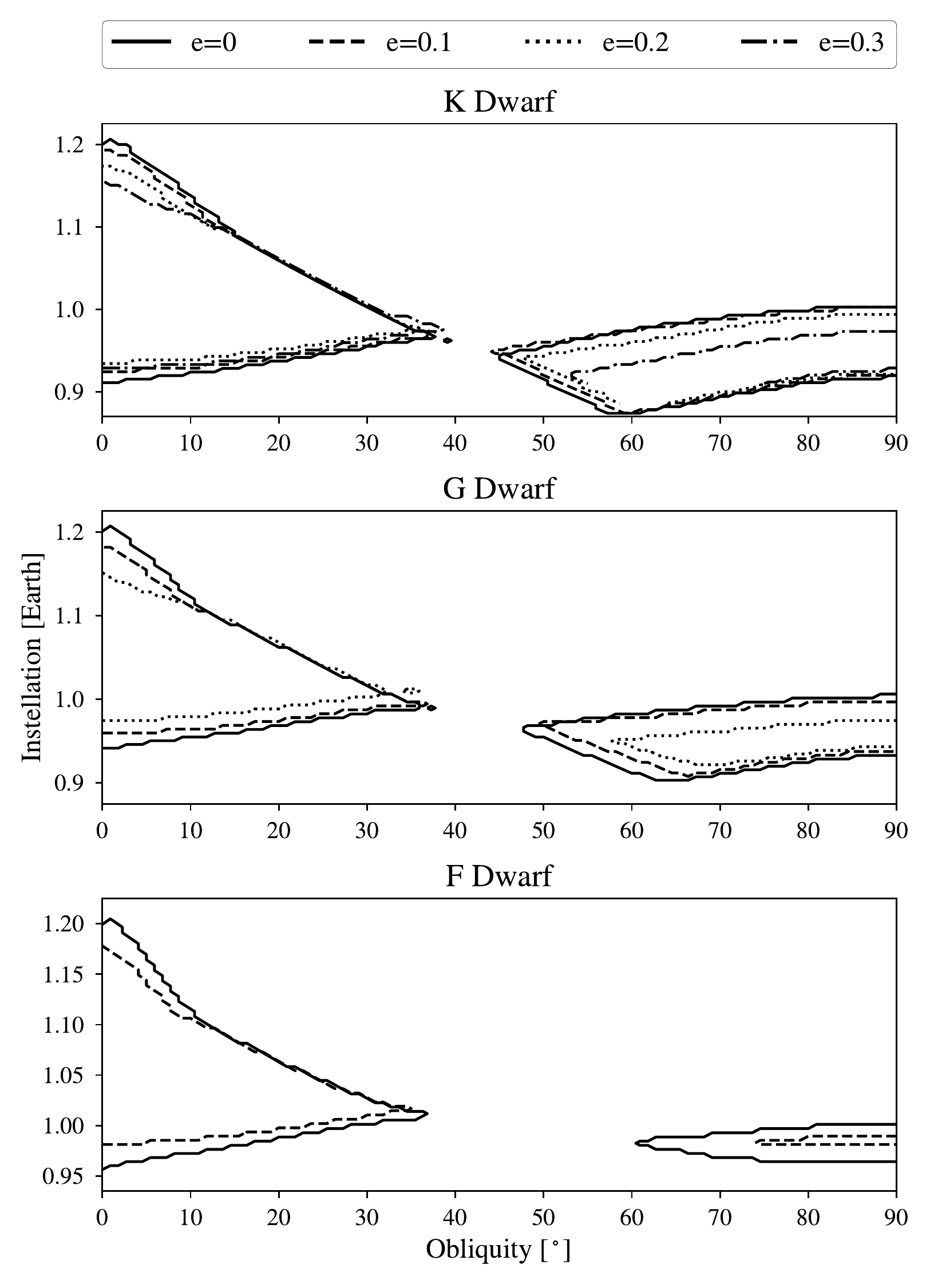}
\caption{Comparisons of ice caps and belt regimes for planets orbiting F (top), G (middle), and K (bottom) as a function of eccentricity, with $e=0$ solid, $e=0.1$ dashed, $e=0.2$ dotted, and $e=0.3$ dash-dotted. Ice belts are unstable at $e>0.3$ for G star planets and $e>0.1$ for F star planets, thus the respective curves are omitted.
\href{https://github.com/caitlyn-wilhelm/IceCoverage/tree/main/EccCompare}{\link{EccCompare}}
}
\label{fig:ecc_compare}
\end{figure}

These results are shown in Fig.~\ref{fig:ecc_compare}, which also shows that ice belts tend to occur at higher $\varepsilon$ as $e$ increases, while the range of $\varepsilon$ is relatively unchanged for ice caps. However, as eccentricity increases, the distribution of ice coverage narrows along the instellation axis---the added effect of eccentricity seasonality makes it harder to stabilize inter-annual ice at the poles or equator. Note that our definition of $S$ is the orbit-averaged instellation, which explicitly includes $e$. Eccentricity increases the orbit-averaged instellation; we thus plot against the orbit-average to place all cases on equal footing. This means that the primary driver of variations shown in Fig. \ref{fig:ecc_compare} is seasonality due to eccentricity, rather than the increase in instellation. 


\subsection{Evolving Planets} \label{sec:evolving}
\begin{figure}[h!]
\centering
\includegraphics[width=0.99\textwidth]{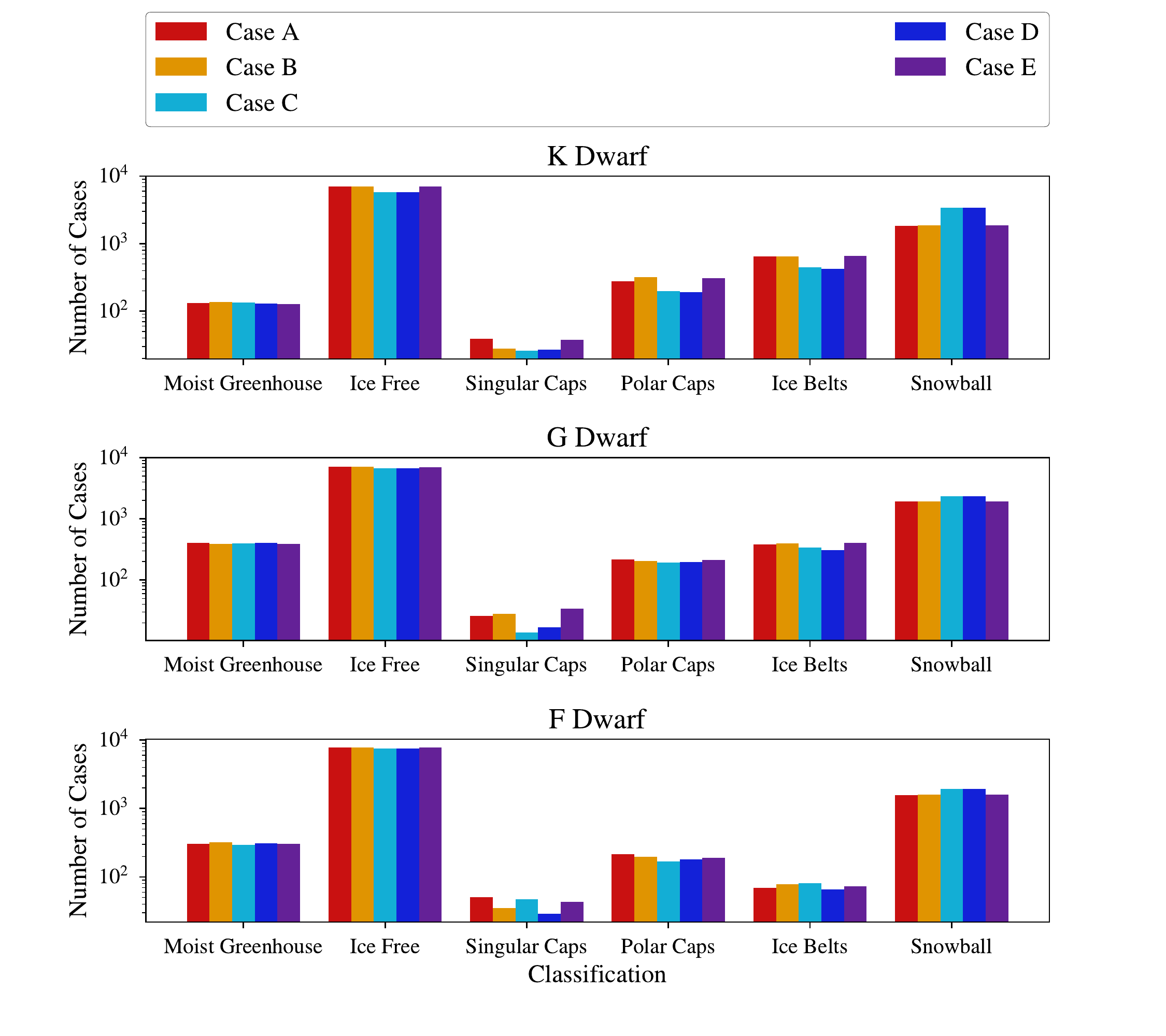}
\caption{Outcomes of our 5 cases of evolving planets, grouped by host star spectral type. (Note the logarthmic scale.) Despite wide ranges of assumptions, the outcomes are relatively constant, suggesting that variations in albedo, rotational cycle, or orbital cycle do not strongly influence the climate states.
\href{https://github.com/caitlyn-wilhelm/IceCoverage/tree/main/DynamicCompare}{\link{DynamicCompare}}
}
\label{fig:albedo_comp}
\end{figure}

Having established the limits of ice states for a range of spectral types and orbits (but with constant albedo), we now turn to planets with evolving orbits and rotational axes, as described in $\S$2.3. We find that ice sheets can persist on many worlds experiencing $e$ and $\varepsilon$ oscillations, in agreement with \cite{Deitrick18b}. We tabulate some statistics of the final climate states of our 150,000 dynamic simulations in Table \ref{albedo_table}, which divides ice belts, polar caps and snowballs into sub categories based on where the ice is present. For polar caps specifically, the mixed column signifies cases in which the two caps possess different types of ice coverage in each , e.g., ice on land in one hemisphere and ice over sea in the other. We visualize the data from Table \ref{albedo_table} in Fig.~\ref{fig:albedo_comp}, which reveals that the vast majority of Earth-like planets orbiting FGK stars are ice free.  If we exclude the simulations that produced a moist greenhouse or snowball state, we predict about 90\% of planets with surface water to be free of surface ice. Crucially, this trend holds for all 5 cases, indicating that this result is robust, at least for our assumptions. The preponderance of ice-free conditions is consistent with Earth's rock record, which indicates that Earth was free from ice caps for at least half of the Phanerozoic Eon \citep{Veizer00}. 

For the remaining 10\%, a few trends stand out. For G and K dwarf planets, we find that ice caps, including unipolar and bipolar, occur with a frequency of about 3--4\%, whereas ice belts occur in about 5--8\% of our cases. Thus, we predict that ice belts are about twice as common as ice caps. This result is primarily due to the initial $\varepsilon$ distribution we invoked, which strongly favors obliquities near $90^\circ$, even though the parameter space for ice caps is larger than that for ice belts, as shown in the top panels of Figs.~\ref{fig:G_example}--\ref{fig:K_star}. As with the static cases, all ice belts exist only on land. As shown below, some cases sustain transient caps and belts, which could bias our results, but here we assume that roughly equal numbers of cases will end in each state and these cases do not skew our conclusions. 

The relative frequency of ice belts and caps for F dwarf planets, however, is reversed, with caps about 3 times more likely than belts. This result is most likely a simple consequence of the relatively smaller size of the ice belt region compared to the polar caps region shown in Fig. \ref{fig:Startype_Compare}. Moreover, the fraction of F star planet cases in which an ice belt was present was below 1\% for all five cases, suggesting that they are very uncommon.

The ratio of the frequency of ice belts to ice caps found by \cite{Rose2017} was about 0.46 when a similar obliquity distribution was assumed. The somewhat higher frequency of ice belts in our study may be a consequence of the thermal inertia of ice sheets and oceans or the time evolution of the orbits and obliquity, neither of which is included in the simpler, analytic model of \cite{Rose2017}. Furthermore, \cite{Rose2017} illustrated that ice belts have more stable real estate in the seasonal model compared to the ice model; our higher frequency of ice belt worlds is likely in part due to our use of the seasonal EBM. Note, however, that the behavior of ice caps is similar in the seasonal and annual models \citep[see Fig. 9 of ][]{Rose2017}. Future work, ideally with 3-D GCMs, could definitively reveal the source of this discrepancy.

Looking at the static cases only, it is tempting to attribute the trends in ice coverage across spectral types (Figs. \ref{fig:G_example} - \ref{fig:ecc_compare}) to the assumed ice albedo. Furthermore, we should naturally expect some variation in the albedo due mixtures of ice types, dust, or meltponds, for example, so results using a single value for each spectral type are probably misleading. \cite{Shields13} use a line-by-line radiative transfer model to compute the albedo of ice assuming different mixtures of snow and blue marine ice, then ran the EBM with each mixture for each spectral type. The values we have chosen in our evolving cases fall roughly in between their 50\% snow and pure snow cases. As seen in Fig. \ref{fig:albedo_comp}, there are small differences in the ice coverage when the ice albedo is allowed to vary. Particularly, extending the upper limit on the ice albedo to 0.75 (Cases C and D) decreases the number of ice free or partially ice covered cases and increases the number of snowball state cases. However, the number of F dwarf ice belt and ice cap cases remains much smaller than for K or G dwarf planets, even when identical albedo distributions are used for all spectral types. All these points indicate that the length of seasons plays the critical role in the stability of surface ice.

\begin{table}[]
\centering
\caption{Number of Simulations in Each Climate State for the Evolving Cases with Varying Albedo}
\begin{footnotesize}
\begin{tabular}{ccccccclcccclccclccc}
\hline
\multirow{2}{*}{Case} & Spectral & Moist & Ice & \multicolumn{3}{c}{Singular Cap} &  & \multicolumn{4}{c}{Polar Cap} &  & \multicolumn{3}{c}{Ice Belt} &  & \multicolumn{3}{c}{Snowball} \\ \cline{5-7} \cline{9-12} \cline{14-16} \cline{18-20} 
 & Type & Greenhouse & Free & Land & Sea & Both &  & Land & Sea & Both & Mixed &  & Land & Sea & Both &  & Land & Sea & Both \\ \hline
\multirow{3}{*}{A} & K & 132 & 7070 & 37 & 2 & 0 &  & 23 & 6 & 231 & 19 &  & 650 & 0 & 0 &  & 14 & 655 & 1160 \\
 & G & 401 & 7054 & 23 & 3 & 0 &  & 9 & 30 & 148 & 29 &  & 385 & 0 & 0 &  & 1 & 1029 & 888 \\
 & F & 307 & 7785 & 33 & 13 & 5 &  & 17 & 51 & 95 & 52 &  & 68 & 0 & 1 &  & 1 & 1043 & 529 \\ \hline
\multirow{3}{*}{B} & K & 138 & 7015 & 23 & 5 & 0 &  & 23 & 15 & 270 & 10 &  & 646 & 0 & 0 &  & 26 & 673 & 1153 \\
 & G & 391 & 7065 & 21 & 7 & 0 &  & 12 & 21 & 134 & 38 &  & 400 & 0 & 0 &  & 1 & 959 & 951 \\
 & F & 319 & 7782 & 27 & 8 & 0 &  & 12 & 35 & 108 & 42 &  & 77 & 1 & 0 &  & 1 & 1039 & 548 \\ \hline
\multirow{3}{*}{C} & K & 134 & 5772 & 26 & 0 & 0 &  & 44 & 2 & 145 & 8 &  & 446 & 0 & 0 &  & 543 & 218 & 2655 \\
 & G & 395 & 6727 & 12 & 2 & 0 &  & 14 & 19 & 134 & 26 &  & 338 & 0 & 0 &  & 92 & 656 & 1579 \\
 & F & 294 & 7483 & 28 & 15 & 4 &  & 6 & 36 & 94 & 32 &  & 81 & 0 & 0 &  & 0 & 855 & 1068 \\ \hline
\multirow{3}{*}{D} & K & 130 & 5773 & 27 & 0 & 0 &  & 42 & 3 & 141 & 7 &  & 424 & 0 & 0 &  & 539 & 249 & 2635 \\
 & G & 406 & 6740 & 12 & 5 & 0 &  & 9 & 34 & 126 & 27 &  & 311 & 0 & 0 &  & 96 & 639 & 1592 \\
 & F & 308 & 7491 & 13 & 13 & 3 &  & 7 & 39 & 90 & 44 &  & 66 & 0 & 0 &  & 2 & 829 & 1093 \\ \hline
 \multirow{3}{*}{E} & K & 128 & 7007 & 28 & 6 & 4 & & 18 & 11 & 259 & 23 & & 660 & 0 & 0 & & 17 & 693 & 1146 \\
 & G & 392 & 7042 & 23 & 10 & 1 & & 7 & 20 & 146 & 39 & & 401 & 0 & 0 & & 0 & 1009 & 910 \\
 & F & 304 & 7795 & 29 & 11 & 3 & & 11 & 30 & 123 & 27 & & 73 & 0 & 0 & & 0 & 1069 & 525 \\ \hline
 \label{albedo_table}
\end{tabular}
\end{footnotesize}
\end{table}

\begin{figure}[h!]
\centering
\includegraphics[width=\textwidth]{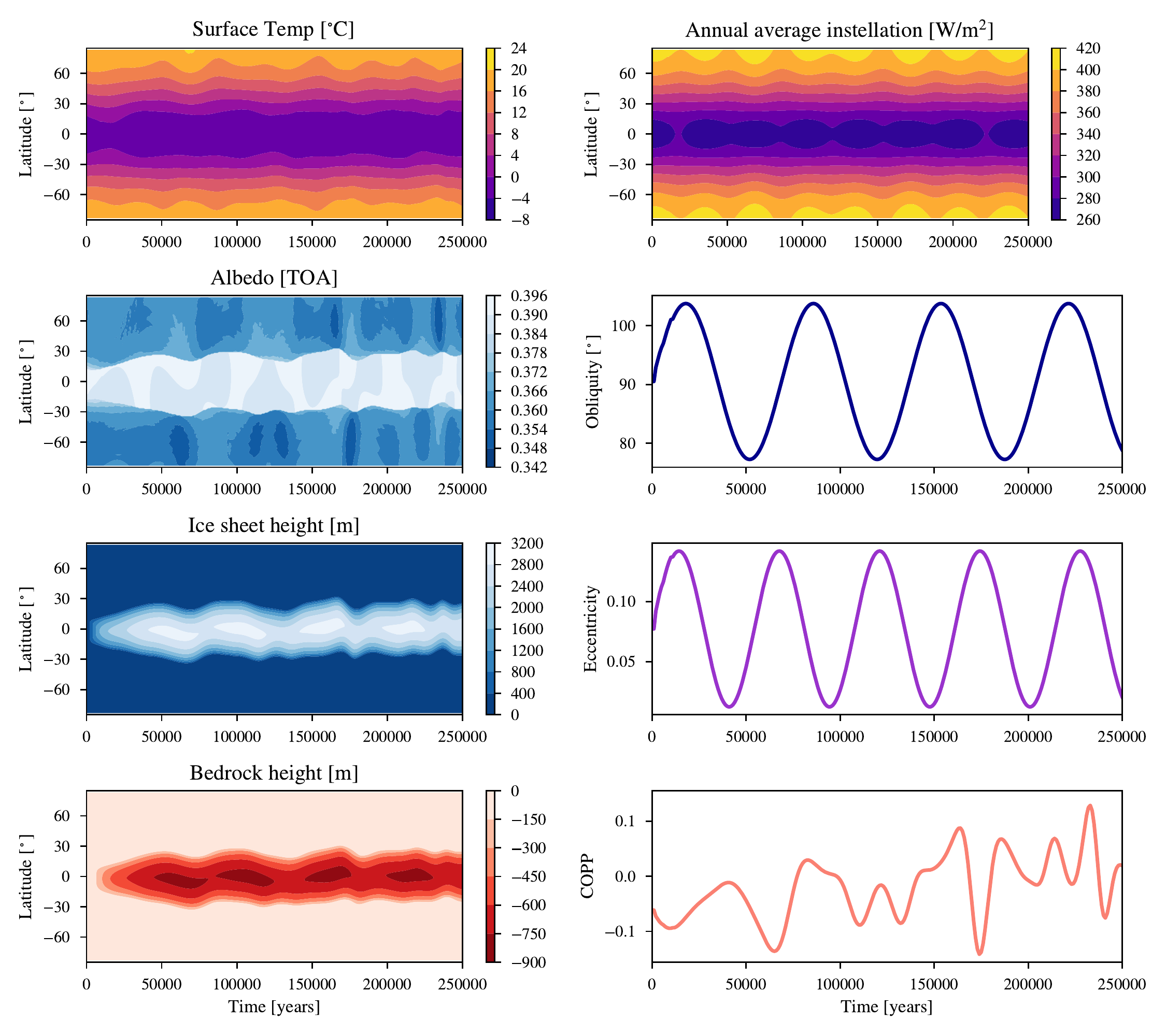}
\caption{An example from the dynamic case G dwarf set that generated a stable ice belt. {\it Top left:} Surface temperature. {\it Top middle left:} Top of atmosphere albedo. {\it Bottom middle left:} Ice sheet height. {\it Bottom left:} Bedrock height (note the negative scale). {\it Top right:} Annual average instellation. {\it Top middle right:} Obliquity.  {\it Bottom middle right:} Eccentricity. {\it Bottom right:} Climate obliquity precession parameter, see Eq.~(\ref{eq:COPP}).
\href{https://github.com/caitlyn-wilhelm/IceCoverage/tree/main/DynamicExample}{\link{DynamicExample}}
}
\label{fig:evolve_example}
\end{figure}

An example case with an ice belt on a planet orbiting a G dwarf star is shown in Fig.~\ref{fig:evolve_example}, with initial conditions presented in Table \ref{tab:select-ic}. The figure shows the ice belt extent does vary due to the oscillations, including excursions into both the northern and southern hemispheres. These variations are correlated with the climate-obliquity-precession parameter (COPP),
\begin{equation}
\chi = e\sin(\varepsilon)\sin(\varpi+\psi),
\label{eq:COPP}
\end{equation}
which describes the relative instellation on the northern and southern hemispheres \citep{Deitrick18b}. This style of Milankovitch cycling also produces maximum lithospheric depression of about 1 km, which is similar to the maximum depth for Earth over the last 2 Myr \citep{Deitrick18b}.

\begin{figure}[h!]
\centering
\includegraphics[width=\textwidth]{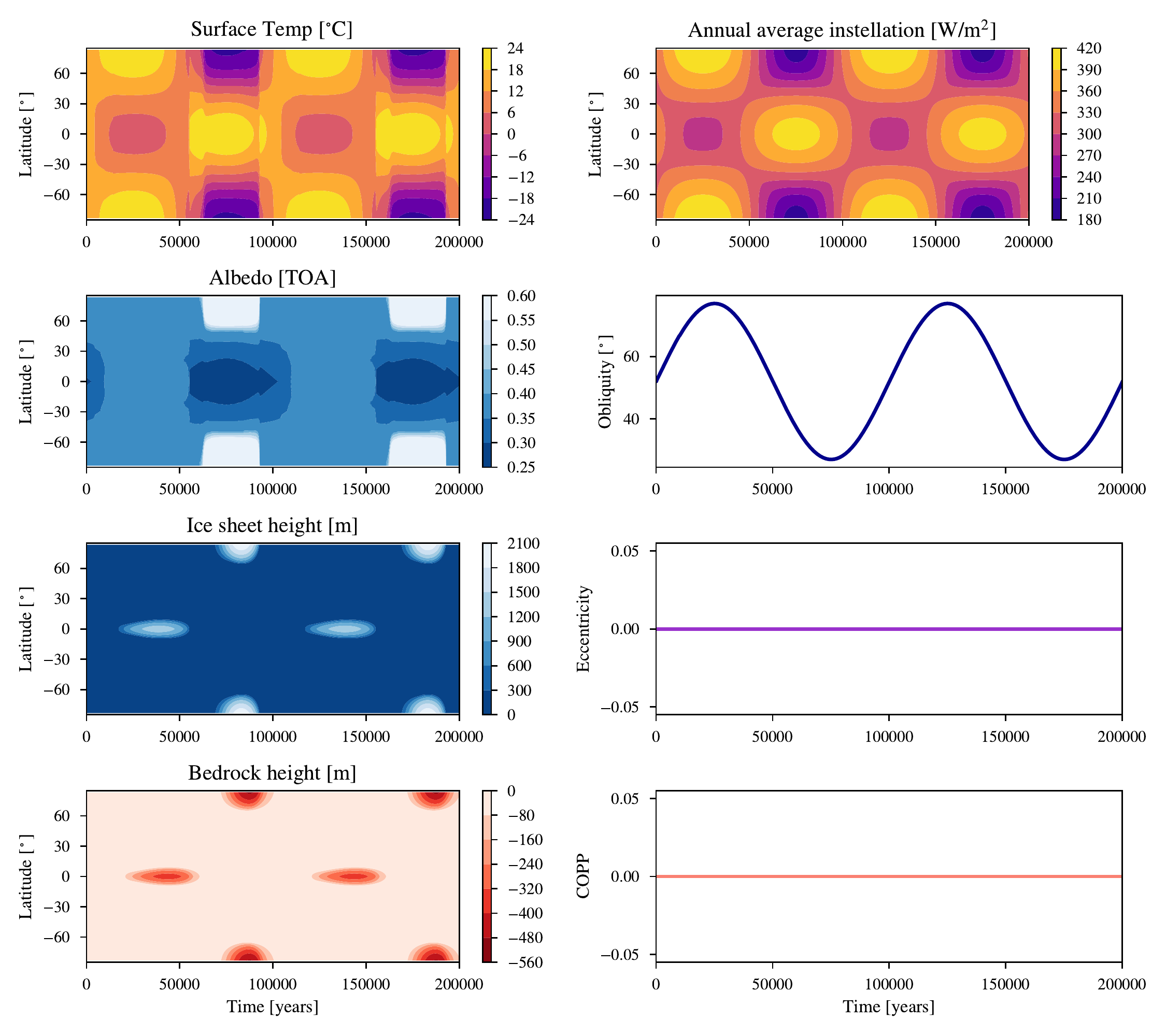}
\caption{An example of a planet whose $\varepsilon$ cycle causes the planet to periodically possess ice belts and ice caps, with ice free epochs in between. The format is the same as Fig.~\ref{fig:evolve_example}.
\href{https://github.com/caitlyn-wilhelm/IceCoverage/tree/main/PeriodicExample}{\link{PeriodicExample}}
}
\label{fig:cycling}
\end{figure}

\begin{figure}[h!]
\centering
\includegraphics[width=\textwidth]{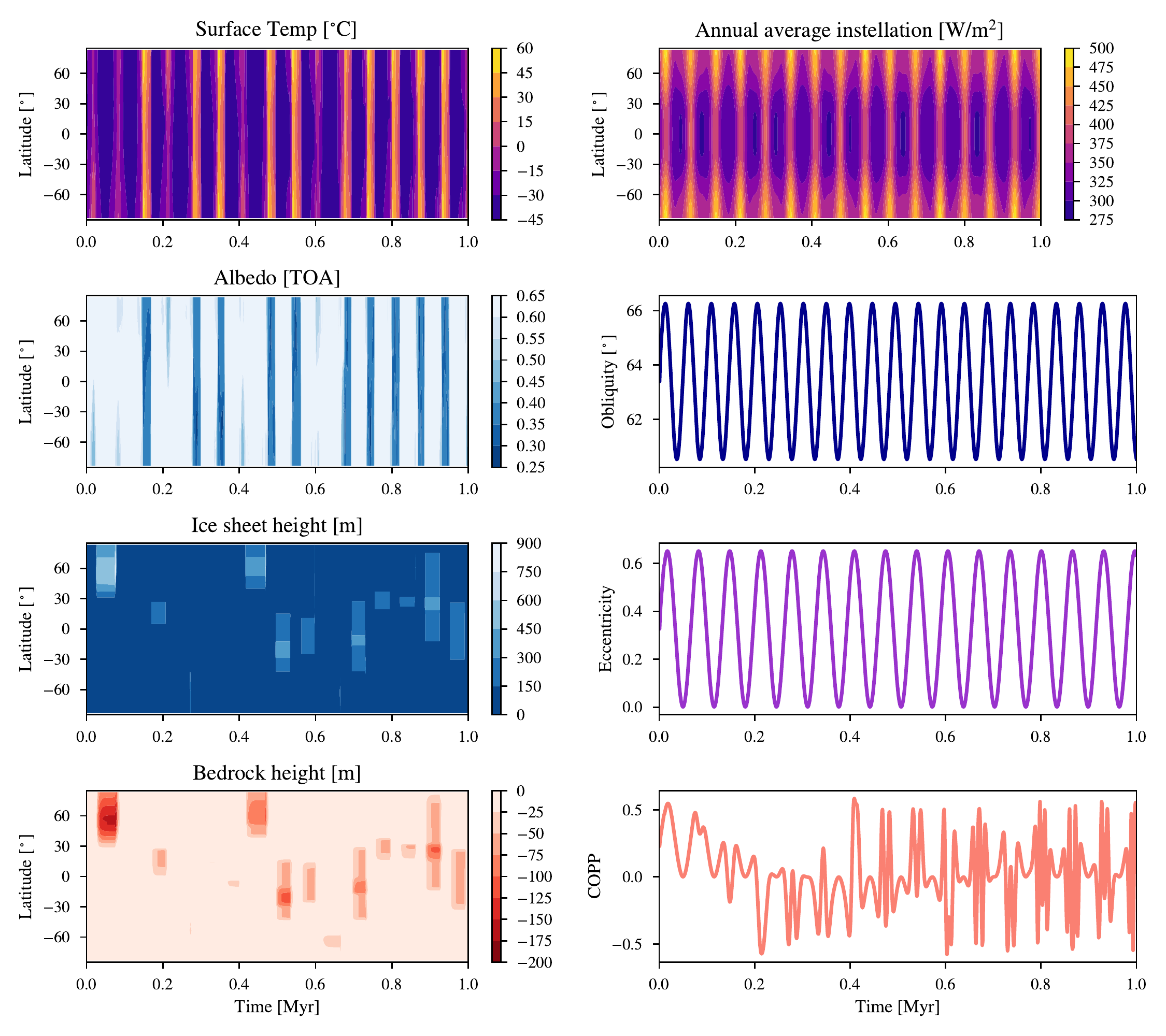}
\caption{An example of a planet whose $\varepsilon$ and eccentricity cycle causes the planet's climate to evolve chaotically, with epochs of both equatorial ice and polar ice. The format is the same as Fig.~\ref{fig:evolve_example}.
\href{https://github.com/caitlyn-wilhelm/IceCoverage/tree/main/ChaoticExample}{\link{ChaoticExample}}
}

\label{fig:chaotic}
\end{figure}

We also found cases in which ice coverage cycles between the cap and equator, either periodically or chaotically. Figure \ref{fig:cycling} shows a periodic case, with initial conditions presented in Table \ref{tab:select-ic}. The initial conditions of this case were tuned to demonstrate this variability in the ice sheets, so while we cannot quantify the frequency of this behavior, it nonetheless would seem likely that some habitable exoplanets can experience this extreme form of Milankovitch cycling. We also found examples in our suite of evolving cases that show chaotic climate evolution, as shown in Fig.~\ref{fig:chaotic}. In this particular case, the planet passes through habitable and uninhabitable surface conditions aperiodically, even though $e$ and $\varepsilon$ evolve sinusoidally. We did not find a single case in which an ice cap and ice belt could stably coexist on a planet. However, there are rare cases ($<1\%$) for which the seas are frozen but the land is not, potentially offering refugia for surface life. These cases tended to occur on relatively cold planets as they briefly exited the snowball state. We note that \cite{Quarles21} independently found that ice coverage can oscillate between the poles and equator.

\begin{table}[]
\centering
\caption{Selected Initial Conditions for Climate Simulations}
\begin{tabular}{cccccccc}
\hline \hline
Figure & Spectral Type & $e_0$ & $P_e$ (kyr) & $A_e$ & $\varepsilon_0$ ($^\circ$) & $P_\varepsilon$ (kyr) & $A_\varepsilon$ ($^\circ$) \\
\hline
\ref{fig:evolve_example} & G & 0.077 & 53326 & 0.13 & 91 & 67823 & 26.5 \\
\ref{fig:cycling} & G  & 0 & N/A & N/A & 52 & 100000 & 50 \\
\ref{fig:chaotic} & F & 0.326 & 65292 & 0.650 & 63.39 & 48192 & 5.743 \\ \hline
\end{tabular}
\label{tab:select-ic}
\end{table}

\begin{figure}[h!]
\centering
\includegraphics[width=\textwidth]{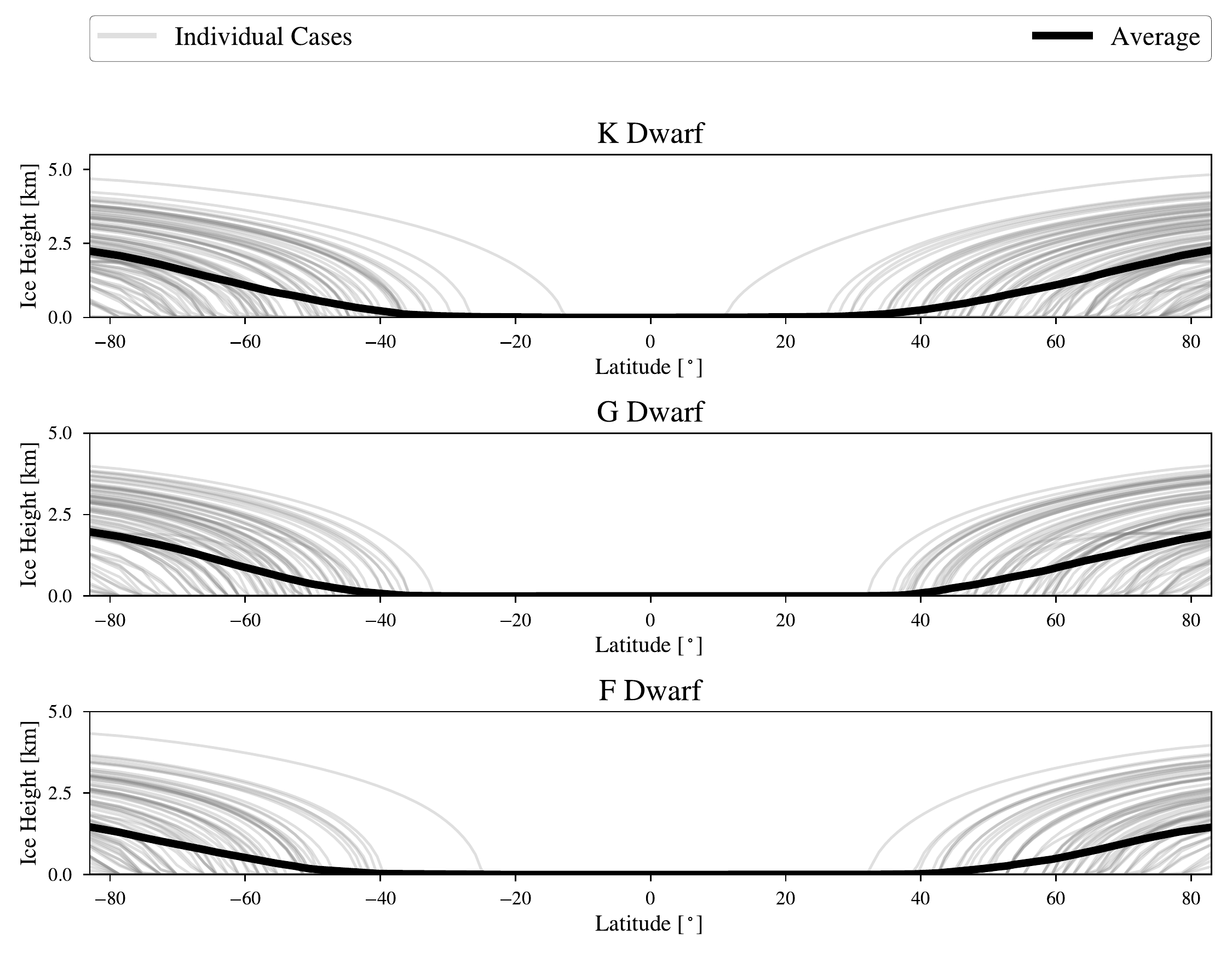}
\caption{Range and average ice heights of ice caps as a function of latitude for stars orbiting F (top), G (middle) and K (bottom) dwarf stars. Note the different scales of the $x$-axes. Light grey curves show 100 randomly selected individual simulations, while black shows the average of all simulations that concluded with an ice belts. Although the averages are all symmetric about the poles, some individual ice caps are significantly displaced. \href{https://github.com/caitlyn-wilhelm/IceCoverage/tree/main/CapHeight}{\link{CapHeight}}
}
\label{fig:cap_height}
\end{figure}

\begin{figure}[h!]
\centering
\includegraphics[width=\textwidth]{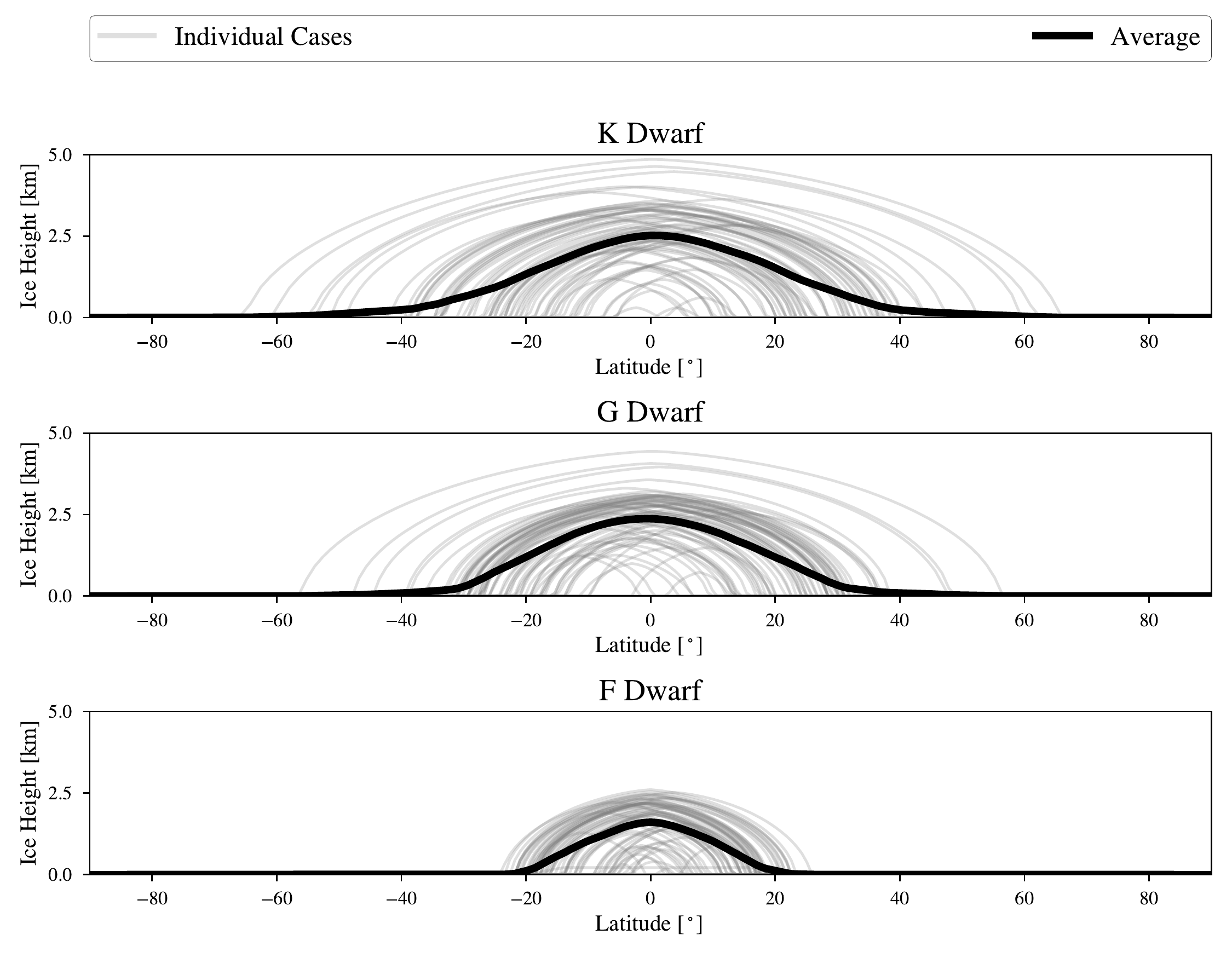}
\caption{Same as Fig.~\ref{fig:cap_height} but for ice belts. Note that although the averages are all symmetric about the equator, some individual ice belts are significantly displaced into either the northern or southern hemisphere. \href{https://github.com/caitlyn-wilhelm/IceCoverage/tree/main/BeltHeight}{\link{BeltHeight}}
}
\label{fig:belt_height}
\end{figure}

Finally, we consider the height and latitudinal extent of the ice sheets in Figs.~\ref{fig:cap_height} and \ref{fig:belt_height}. These figures show the ice height as a function of latitude at the end of all simulations that possessed ice caps or an ice belt, respectively. Individual cases are shown in light grey, with the mean ice height shown in black. In general, the cooler the host star, the larger and taller the ice sheets. Most ice belts span the equator, but are not symmetric about it. However, some ice belts are centered more than $10^\circ$ from the equator, see \eg Fig.~\ref{fig:chaotic}. These asymmetric belts can occur when the COPP value deviates significantly from 0, \ie one hemisphere is much cooler than the other, \cf Fig.~\ref{fig:evolve_example}. Thus, the extent of the ice caps and position of the ice belts appears to correlate with $\chi$, see Eq.~(\ref{eq:COPP}). 

\section{Conclusions}
We have performed a suite of simulations that suggest a) Earth-like planets are very likely to be ice free, b) ice belts are stable on land for planets orbiting FGK dwarfs stars, even when experiencing eccentricity and obliquity oscillations, and c) ice belts are about twice as common as polar ice caps for planets orbiting G and K dwarf stars. These results may be used to plan future direct imaging observations of potentially habitable exoplanets. 

Our experiments elucidated numerous features about ice belts that were previously unknown. We find belts are much more likely to be found on planets orbiting K and G dwarf stars than F dwarf stars since a) the albedo of water ice is larger for F dwarf spectra, increasing the likelihood of collapse into a snowball and  b) the longer winters allow the ice belts to grow to a critical state that triggers the snowball instability. We find that the typical extent of an ice belt is 10$^\circ$ -- 30$^\circ$ depending on host star spectral type, potentially large enough to imprint a signal on photometric and spectroscopic measurements. Thus, future surveys of habitable planets of K and G dwarfs should prepare for the possibility that they will detect a world with an ice belt. 

We found that cold start planets are unlikely to form polar caps, see Figs.~\ref{fig:G_example}--\ref{fig:K_star}, suggesting that any exoplanet discovered with them has recently experienced an ice free state. This result also provides further support for the theory that as Earth exited Snowball Earth phases, it possessed enough CO$_2$ to trigger a ``hot house'' climate that completely melted surface ice \citep{Kirschvink92,Hoffman99}. Only after CO$_2$ levels dropped and stabilized could polar caps form.

Our dynamic cases highlight the importance of considering currently ice-covered planets as potentially habitable because they may have recently possessed open surface water. Such worlds could still develop life in a manner similar to Earth, e.g. in wet/dry cycles on land, but then the dynamics of the planetary system force the planet into a snowball, which in turn forces life into the ocean under a solid ice surface. Such a process may have transpired multiple times on Earth, so we should expect similar processes to function on exoplanets.

While previous work \citep{WilliamsPollard03,Rose2017} found that the critical $\varepsilon$ for ice belt formation was about 55$^\circ$, we find that in most cases the critical values are smaller. Those previous studies used annual EBMs, whereas we used a seasonal model, which is the likely origin of the discrepancy. In Figures 9 and 10 of Rose et al, they compare their results from the analytic annual model with the numerically integrated seasonal model. While these compare fairly well at low $\varepsilon$, there are large discrepancies at high $\varepsilon$, \ie where ice belts form. In conjunction with our results, it appears that the annual model is too simple to accurately derive the critical $\varepsilon$.

While we have made quantitative predictions about ice state frequencies, we must also acknowledge several approximations in our model that could affect our results. First, our EBM is idealized and does not take into account the 3-D nature of a planet, including zonal winds and cloud variability. Furthermore, we assumed a uniform land/sea ratio as a function of latitude; different configurations will assuredly lead to different results \citep[\cf][]{WilliamsPollard03,Rushby19}. Second, we assumed the diffusion coefficients of the EBM \citep[see][]{Deitrick18b,Barnes20} are constant regardless of $\varepsilon$ and $e$, which may not be appropriate. We calibrated our EBM to modern Earth, but meridional heat flow is likely a function of $\varepsilon$. Third, we do not include ocean heat transport, which \cite{Ferreira14} found strongly influenced ice belt stability. Fourth, our oscillations are idealized and do not necessarily reflect actual planet-planet perturbations. Fifth, we did not include geochemical processes such as the carbonate-silicate cycle \citep{Walker81}. These phenomena can significantly affect the radiative properties of an atmosphere, \eg greenhouse warming, and lead to outcomes that are not described here, such as climate limit cycles \citep{HaqqMisra16,Abbot16}. Each of these assumptions was made in the name of tractability as current computational software and hardware limitations prevent the broad parameter sweeps presented here to include these physics and still be completed in a reasonable amount of wallclock time. Future research that addresses these deficiencies could modify the results presented above.

In this study we did not consider planets orbiting M dwarfs as described in $\S$ 1. In single planet systems, tidal torques are likely to drive $\varepsilon$ near 0 or $\pi$ \citep{Heller11}. However, in multiplanet systems, $\varepsilon$ can settle into an approximately constant non-zero value \citep{Dobrovolskis09,Barnes20} with the planet in synchronous rotation. Future simulations could explore the frequency and locations of ice sheets on M dwarf planets with a GCM to determine if the results presented here extend to systems with lower mass host stars. Such methods could also be applied to habitable planets orbiting brown dwarfs or white dwarfs \citep{Agol11,Bolmont12,BarnesHeller13}. 

Finally, we can use the results of $\S$3 to estimate the frequency of ice types of FGK habitable planets from our 4 evolving planet cases. For this calculation we remove the planets that ended in a moist greenhouse or snowball state as they are, by definition, uninhabitable. After summing the classifications from all 5 cases, we then have 39,858 habitable F dwarf planets, 37,604 habitable G dwarf planets, and 36,921 habitable K dwarf planets in our sample. For G dwarf planets, the ice state frequencies are 92\% ice free, 2.7\% polar cap(s), and 4.8\% ice belt. For F dwarf planets, the percentages are 96.1\%, 2.9\%, and 0.9\%, respectively. For K dwarf planets, the percentages are 88.4\%, 3.5\%, and 7.6\%, respectively. Thus, we predict the vast majority of habitable Earth-like planets of FGK stars will be ice free, that ice belts will be twice as common as caps for G and K dwarfs planets, and that ice caps will be three times as common as belts for Earth-like planets of F dwarfs.

Our conclusions may be testable with future 10-m class space telescopes that may be capable of detecting surface features of planets orbiting in the HZs of FGK stars \citep{CowanFujii18,Luger21}. Similarly, 30-m class ground telescopes may be able to map the surfaces of habitable planets orbiting early M dwarfs \citep{Skidmore15,Rodler18}, which may require billions of years to tidally lock \citep{Barnes17}. Simulations like those described here can serve as inputs to instrument simulators to predict the photometric and/or spectroscopic signal of different ice states. Such synthetic observations may prove more fruitful with GCMs, but at least our results can guide those computationally expensive experiments to critical locations in parameter space. Future research on ice coverage could provide critical insight into the design of large-aperture telescopes, the preparation of retrieval methods, and, possibly, the first discovery of a habitable exoplanet.

\medskip

This work was supported by NASA's Virtual Planetary Laboratory under cooperative agreements NNA13AA93A and 80NSSC18K0829. We thank Cecilia Bitz for valuable discussions and guidance on this research. We also thank Daniel Koll and an anonymous referee for comments that greatly improved the clarity and scope of this manuscript.

\bibliography{references}
    
\end{document}